\documentclass[12pt]{article}
\usepackage{amsmath}
\usepackage{graphicx}
\usepackage{enumerate}
\usepackage{natbib}
\usepackage{booktabs}
\usepackage{bm}
\usepackage{url,xcolor}
\usepackage[linesnumbered,ruled,vlined]{algorithm2e}
\usepackage{amssymb,amsthm}
\usepackage{geometry,setspace}
\usepackage{pdfpages}
\newtheorem{theorem}{Theorem}

\newtheorem{lemma}{Lemma}
\newtheorem{condition}{Condition}

\newcommand{\bX}{\boldsymbol{X}}
\newcommand{\bZ}{\boldsymbol{Z}}
\newcommand{\bI}{\boldsymbol{I}}
\newcommand{\bbeta}{\boldsymbol{\beta}}
\newcommand{\cf}{\widehat{f}}
\newcommand{\btheta}{\boldsymbol{\theta}}
\newcommand{\hbtheta}{\widehat{\boldsymbol{\theta}}}
\newcommand{\tbtheta}{\widetilde{\boldsymbol{\theta}}}
\newcommand{\htheta}{\widehat{\bm\theta}}

\newcommand{\bv}{\boldsymbol{v}}
\newcommand{\bh}{\boldsymbol{h}}
\newcommand{\be}{\boldsymbol{e}}
\DeclareMathOperator*{\argmax}{arg\,max}
\DeclareMathOperator*{\argmin}{arg\,min}
\newcommand{\pr}{\text{Pr}}

\title{Debiased Inference for High-Dimensional Regression Models Based on Profile M-Estimation}

\author{Yuhao Deng$^*$, Yi Wang\thanks{Yuhao Deng and Yi Wang contributed equally.}, Yu Gu, Yuanjia Wang, Donglin Zeng}

\date{}

\begin{document}

\maketitle

\begin{abstract}
Debiased inference for high-dimensional regression models has received substantial recent attention to ensure regularized estimators have valid inference. Many existing methods focus on achieving Neyman orthogonality through explicitly constructing projections onto the space of nuisance parameters, which is infeasible when an explicit form of the projection is unavailable.
We introduce a general debiasing framework, Debiased Profile $M$-Estimation (DPME), which applies to a broad class of models and does not require model-specific Neyman orthogonalization or projection derivations as in existing methods. Our approach begins with obtaining an initial estimator of the parameters by optimizing a penalized objective function. To correct for the bias introduced by penalization, we construct a one-step estimator using the Newton--Raphson update, applied to the gradient of a profile function defined as the optimal objective function with the parameter of interest held fixed. We use numerical differentiation without requiring explicit calculation of the gradients.
The resulting DPME estimator is shown to be asymptotically linear and normally distributed. Through extensive simulations, we demonstrate that the proposed method achieves better coverage rates than existing alternatives with largely reduced computational cost. Finally, we illustrate the utility of our method by applying it to estimate a treatment rule for multiple myeloma.
\end{abstract}


\section{Introduction}
\label{sec:intro}

Statistical inference in high-dimensional regression models, where the number of predictors can substantially exceed the sample size, poses significant challenges. Regularization is essential for estimation, but it introduces non-negligible bias into the resulting estimators, which consequently lack the desirable asymptotic properties, such as $\sqrt{n}$-rate convergence and asymptotic normality—enjoyed in low-dimensional settings. 
For example, it is well known that the Lasso estimator is not $\sqrt{n}$-consistent \citep{candes2007dantzig, zhang2008sparsity, buhlmann2011statistics}, and traditional bootstrap methods do not provide valid statistical inference in this case \citep{chatterjee2010asymptotic}.
To overcome these challenges, debiasing methods have received considerable attention in recent years, and they have been successfully applied to a wide range of models and problems, including high-dimensional linear or generalized linear models \citep{zhang2014confidence, van2014asymptotically, ning2017general}, high-dimensional Gaussian graphical models \citep{ren2015asymptotic}, high-dimensional additive models \citep{gregory2021statistical}, high-dimensional single-index models \citep{eftekhari2021inference}, and treatment effect estimation \citep{chernozhukov2022automaticdynamic}.

The central idea of debiasing methods is to correct the bias of an initial estimator by constructing a corrected score function for the parameter of interest that is orthogonal to the nuisance score, a property known as Neyman orthogonality \citep{foster2023orthogonal}. Neyman orthogonality ensures that the resulting estimator is insensitive to first-order perturbations in the nuisance parameters, enabling $\sqrt{n}$-consistent and asymptotically normal inference even when the nuisance parameters are estimated at slower rates. This property is central not only to classical debiasing methods but also to a broader class of debiased machine learning methods. For example, the decorrelation method introduced by \citet{ning2017general} constructs a score function uncorrelated with the nuisance score in closed form and estimates it via a Dantzig-type estimator;  \citet{chernozhukov2018double} and \citet{mackey2018orthogonal} employed Neyman orthogonal scores for two-stage treatment effect estimation in partially linear regression models; \citet{chernozhukov2022automaticcausal} studied debiased estimation of causal and structural effects via an automatic and nonparametric construction of the orthogonal score; \citet{farrell2020deep} proposed a two-stage automatic inference framework in which a machine-learning-enriched model is carried forward to a low-dimensional parameter of interest based on the Neyman orthogonal score construction.
More recently, \citet{van2025automatic} developed an automatic debiasing method for smooth functionals of nonparametric $M$-estimands by expressing the estimation error via the Hessian Riesz representation and automatically estimating the representor via a risk minimization problem. 
However, many existing debiasing methods require constructing these projections or the Neyman orthogonal score on a model-specific, case-by-case basis. Doing so requires detailed analytical knowledge of the model structure and its parameters, as well as the ability to derive projections onto specific high-dimensional spaces. Moreover, these methods generally assume that the parameter of interest and the nuisance parameters enter the model in a separable fashion, and they rely on the explicit construction of a loss function that satisfies Neyman orthogonality. In many practical settings, such closed-form projections do not exist or are intractable to derive analytically. Identifying the Neyman orthogonal score may then require numerically solving high-dimensional systems of equations, making computation unstable and prohibitively expensive. 

To avoid the burden of analytically deriving Neyman orthogonal scores, an alternative line of work has adopted the idea of profiling \citep{murphy1997maximum, murphy2000profile}, in which the influence of nuisance parameters is eliminated by treating them as implicit functions of the parameter of interest. The profile objective function is obtained by optimizing out the nuisance parameters at each fixed value of the target parameter, so that it depends only on the parameter of interest. The influence of nuisance parameters on the inference of the target parameter is thus eliminated, yielding Neyman orthogonality via this numerical treatment. With this objective function as a function of only the parameter of interest, inference can proceed directly on this reduced function. Inspired by the idea of profiling, we propose a general debiasing framework, referred to as the Debiased Profile $M$-Estimation (DPME). Our framework accommodates penalized regression and does not require any knowledge of how to derive projections, decorrelated scores, or Neyman orthogonal scores. The method begins by obtaining an initial estimator of the parameters by optimizing a penalized objective function, which naturally accommodates high dimensionality. To correct the bias introduced by penalization, we construct a one-step estimator using the Newton--Raphson update applied to the gradient of a profile function. The profile objective function is defined as the minimum of the loss over the nuisance parameters at each fixed value of the parameter of interest. For computational feasibility, we use numerical differentiation to approximate the gradient and Hessian of the profile objective function, eliminating the need for any explicit analytic derivation of projections or orthogonal scores. This strategy makes DPME broadly applicable and computationally efficient across a wide range of high-dimensional models.

Our DPME framework offers several important advantages. 
First, our debiasing procedure only requires calculating numerical derivatives of the profile objective function with respect to the low-dimensional parameter of interest, avoiding any explicit analytic derivation of orthogonal scores or high-dimensional projections.
Second, computation in our approach reduces to repeated penalized $M$-estimation, eliminating high-dimensional projections and thus greatly lowering computational complexity. 
Third, the procedure does not require access to the internal details of the specific estimation methods employed. That is, as long as one can evaluate the penalized objective function, the DPME correction can be implemented as a black-box procedure, making it readily automatable for a broad class of $M$-estimation problems. We establish the asymptotic normality and $\sqrt{n}$-consistency of the proposed DPME estimator under regularity conditions, and demonstrate its finite-sample performance through extensive simulations and real data applications.

The rest of this paper is organized as follows. In Section \ref{sec:Method}, we describe the step-by-step details of DPME. In Section \ref{sec:asymptotic_results}, we provide a list of regularity conditions to show that the estimators from DPME are asymptotically linear and normally distributed. In Section \ref{examples}, we apply the proposed method to study high-dimensional Lasso, generalized linear model, optimal treatment rules, and verify all necessary conditions for these examples. Section \ref{simulation} presents simulation results from extensive simulation studies that use these models as examples. Finally, in Section \ref{application}, we apply our approach to learn optimal linear decision rules for treating multiple myeloma.

\section{Methods}
\label{sec:Method}

Consider a random sample of $n$ independent copies of $\bZ^{(n)} \sim P^{(n)}$, where $\bZ^{(n)}$ has $p_n$ dimensions and $P^{(n)}$ is a probability measure on $\mathbb{R}^{p_n}$. 
Let $f_n$ denote the complete functional parameter associated with $P^{(n)}$, which can be high-dimensional and is typically estimated via penalized $M$-estimation. 
We are only interested in a $q$-dimensional functional ($q$ is fixed) of $f_n$, denoted by $\btheta_{n} = \mathfrak{F}_n(f_{n})$.  
For simplicity of presentation, we suppress the superscript ``$(n)$'' in $\bZ^{(n)}$ and $P^{(n)}$, as well as the subscript ``$n$'' in $f_n$, $\btheta_n$, and $\mathfrak{F}_n$, with the understanding that all these terms may depend on the sample size $n$.
To further understand the notation, we use high-dimensional linear regression as a simple illustrative example.
The data is structured as $\bZ = (\bX, Y)$, where $\bX$ represents the high-dimensional covariates and $Y$ is the response variable. 
Consider the linear model $Y = \bX^{\top} \bbeta + \epsilon$, where $\bbeta$ is a vector of regression coefficients and $\epsilon$ is a random error term. 
Here, the complete parameter $f$ is $\bbeta$, and the parameter of interest $\btheta$ can be a subvector of $\bbeta$, such as its first element.
The functional $\mathfrak{F}$ simply maps $\bbeta$ to its corresponding subvector.

Assume that the true value of $f$, denoted by $f_{0}$, is the unique maximizer of $P m(\bZ, f)$ among all $f\in \mathcal{F}$, where $P$ is the true probability measure, $\mathcal{F}$ is a known Hilbert space equipped with an inner product $\langle \cdot, \cdot \rangle$, and $m: \mathbb{R}^{p_n}\times \mathcal{F}\rightarrow \mathbb{R}$ is some objective function.
A typical way to estimate $f$ in high-dimensional scenarios is via penalized $M$-estimation:
\begin{equation} \label{initial_estimator}
\widehat{f}_n = \argmax_{f\in\mathcal{F}} \mathbb{P}_n m(\bZ, f) - \lambda_n J(f),
\end{equation}
where $\mathbb{P}_n$ is the empirical probability measure based on $n$ observations, $J(\cdot)$ is a prespecified penalty function, and $\lambda_n>0$ is a tuning parameter that controls the sparsity or smoothness of $\cf_n$.
Then, an initial estimator of $\btheta$ is given by $\hbtheta_n = \mathfrak{F}(\cf_n)$.

In high-dimensional settings, penalization is necessary to prevent the variance of the estimator from diverging. 
However, penalization also introduces substantial bias to $\widehat\btheta_n$, making it irregular and not $\sqrt{n}$-consistent. 
To enable valid statistical inference on $\btheta$, it is important to remove the effects of penalization by correcting the bias in $\widehat\btheta_n$.
%This can be accomplished through a simple one-step update, which we introduce next.
To this end, for every fixed $\btheta$, let $\cf_n(\btheta)$ denote the estimator of $f$ under the constraint that $\mathfrak{F}(f) = \btheta$, that is,
\[
  \cf_n(\btheta) = \argmax_{f\in\mathcal{F},\, \mathfrak{F}(f) = \btheta} \mathbb{P}_n m(\bZ, f) - \lambda_n J(f).
\]
In many common high-dimensional settings, particularly when $\btheta$ is well separated from the nuisance parameters, this constrained optimization problem can be easily solved.
For example, in high-dimensional linear regression, if $\btheta$ is a subvector of the regression coefficients, solving the above problem reduces to fitting another penalized regression by fixing $\btheta$; see Section~\ref{examples} for details. 
We define $A_n(\btheta) = \mathbb{P}_n m\{\bZ,\cf_n(\btheta)\}$ and refer to $A_n(\btheta)$ as the profile $m$-function, in the sense that the nuisance parameters have been profiled out. 

Our proposed DPME estimator for $\btheta$ is obtained via a one-step update of the initial estimator $\hbtheta_n$, using numerical derivatives of the profile $m$-function.
Specifically, let $D_{h_{1n}}$ and $D^2_{h_{2n}}$ denote the first- and second-order numerical derivatives with sufficiently small perturbation constants $h_{1n}$ and $h_{2n}$, respectively. 
Suppose that $\btheta$ has $q$ dimensions. 
For the first-order derivative, we evaluate the $j$th element of $D_{h_{1n}}A_n(\btheta)$ by  
\begin{equation}
  \frac{A_n(\btheta + \be_jh_{1n}) - A_n(\btheta - \be_jh_{1n})}{2h_{1n}}, \label{D1_cal}
\end{equation}
where $\be_j$ is the $j$th canonical vector in $\mathbb{R}^q$, for $j \in \{1,\ldots,q\}$.
For the second-order derivative, we evaluate the $(j,k)$th element of $D^2_{h_{2n}}A_{n}(\btheta)$ by 
\begin{equation}
  \frac{A_{n}(\btheta+\be_jh_{2n}+\be_kh_{2n}) - A_{n}(\btheta+\be_jh_{2n}-\be_kh_{2n}) - A_{n}(\btheta-\be_jh_{2n}+\be_kh_{2n}) + A_{n}(\btheta-\be_jh_{2n}-\be_kh_{2n})}{4h_{2n}^2}, \label{D2_cal}
\end{equation}
for $j,k \in \{1,\dots,q\}$.
Then, we debias the initial estimator $\hbtheta_n$ via a one-step update:
\begin{equation} \label{debias}
  \tbtheta_n = \hbtheta_n - \left\{D^2_{h_{2n}}A_{n}(\widehat\btheta_n)\right\}^{-1} D_{h_{1n}}A_n(\widehat\btheta_n).
\end{equation}

%Figure \ref{fig1} provides a graphical illustration of this debiasing step: the initial estimator $\hbtheta_n$ may be substantially biased due to penalization, but the one-step update shifts it to $\tbtheta_n$, which is closer to the true parameter value $\btheta_0 = \mathfrak{F}(f_0)$.
%\begin{figure}
%    \centering
%    \includegraphics[width=0.6\textwidth]{graph2}
%    \caption{A graphical illustration of the debiased profile $M$-estimation.} \label{fig1}
%\end{figure}

The rationale of the one-step update in \eqref{debias} for bias correction is as follows. 
First, $D_{h_{1n}}A_n(\btheta)$ serves as a numeric approximation to the gradient of the non-penalized objective function. 
Thus, the bias correction term in \eqref{debias} mitigates the impact of penalization, moving the initial estimator $\hbtheta_n$ closer to the maximizer of the non-penalized objective function.
Second, as rigorously established in Section~\ref{sec:asymptotic_results}, the gradient of the profile $m$-function essentially lies in the orthogonal complement of the tangent space for nuisance parameters, thus is insensitive to the estimation error of these nuisance parameters. 
Therefore, using numerical derivatives of the profile $m$-function effectively reduces the bias of $\hbtheta_n$.
By employing numerical derivatives in the one-step update based on Equation \eqref{debias}, our approach circumvents the need for explicit expressions of the first- and second-order derivatives of $A_n(\btheta)$, and is thus broadly applicable to a wide range of high-dimensional problems. 

As a remark, our idea of a one-step update can also be seen in some earlier work on targeted maximum likelihood estimation, where the influence function for the target parameter is corrected at the initial estimator \citep{van2006targeted, van2016one, hines2022demystifying}. Additionally, the one-step debiased estimator $\tbtheta_n$ is closely related to the one-step Newton--Raphson solution to the efficient score function in semiparametric literature, with $D_{h_{1n}}A_n(\btheta)$ playing a role analogous to the efficient score for $\btheta$ that is proven to be Neyman orthogonal to the influence of other parameters. However, existing targeted learning methods require the explicit derivation of the efficient score when the model is not fully nonparametric \citep{van2025automatic, luedtke2026simplifying}, which we avoided by using numerical differentiation for constrained profiled functions.

Under certain regularity conditions, the bias correction term $\{D^2_{h_{2n}}A_{n}(\widehat\btheta_n)\}^{-1} D_{h_{1n}}A_n(\widehat\btheta_n)$ can be shown to provide an asymptotically linear estimator for the bias of  $\hbtheta_n$. 
Thus, the one-step debiased estimator $\tbtheta_n$ is regular and asymptotically linear.
Leveraging the asymptotic linear expansion established in Section~\ref{sec:asymptotic_results}, the asymptotic variance of $\tbtheta_n$ can be consistently estimated by the following sandwich estimator: 
\begin{equation} \label{variance_estimator}
\widehat{\mbox{var}}(\tbtheta_n) = n^{-1} \left\{D^2_{h_{2n}}A_{n}(\tbtheta_n)\right\}^{-1} \mathbb{P}_n \left[D_{h_{1n}}m\{\bZ,\cf_n(\tbtheta_n)\}\right]^2 \left\{D^2_{h_{2n}}A_{n}(\tbtheta_n)\right\}^{-1}.
\end{equation}
Note that the perturbation constants $h_{1n}$ and $h_{2n}$ used for variance estimation can differ from those used in \eqref{debias}.

The complete DPME procedure is outlined in Algorithm~\ref{algo:ppme}. First, an initial estimator $\hbtheta_n$ is obtained from penalized regression. Second, we perturb $\hbtheta_n$ and evaluate the profile $m$-function by penalized regression. Third, we calculate first- and second-order numerical derivatives of the profile $m$-function. Fourth, we debias $\hbtheta_n$ through a one-step update to get the final estimator $\tbtheta_n$. Lastly, we estimate the variance using the sandwich estimator by perturbing $\tbtheta_n$ similarly.
The algorithm simply involves a few rounds of optimization over $f$ and the numerical computation of derivatives of the profile $m$-function with respect to $\btheta$, and thus is computationally efficient.
Importantly, the DPME procedure is highly flexible and does not require detailed knowledge of the method used to estimate $f$ in practice.

%\begingroup
%\renewcommand{\baselinestretch}{1}\normalsize
\begin{algorithm}[!htb]
\SetAlgoLined
\KwIn{Data $\{\bZ_i: i=1,\dots,n\}$, perturbation constants $h_{1n}$ and $h_{2n}$}
\KwOut{Debiased estimator $\tbtheta_n$ and its variance estimator}
 Obtain initial estimators of $f$ and $\btheta$ (with $\lambda_n$ chosen by cross-validation):
 \[
 \widehat{f}_n \leftarrow \argmax_{f\in\mathcal{F}} \mathbb{P}_n m(\bZ, f) - \lambda_n J(f), \quad \widehat\btheta_n \leftarrow \mathfrak{F}(\widehat{f}_n), \quad A_n(\widehat\btheta_n) \leftarrow \mathbb{P}_n m(\bZ,\widehat{f}_n)
 \] 
 
 \For{$\btheta\in\{\widehat\btheta_n \pm \be_kh_{1n}, \widehat\btheta_n \pm \be_jh_{2n} \pm \be_kh_{2n}: j,k=1,...,q\}$}{
 \[
  \cf_n(\btheta) \leftarrow \argmax_{f\in\mathcal{F},\, \mathfrak{F}(f) = \btheta} \mathbb{P}_n m(\bZ, f) - \lambda_n J(f),
  \quad A_n(\btheta) \leftarrow \mathbb{P}_n m\{\bZ,\cf_n(\btheta)\}
  \]
 }
 %Compute $D_{h_{1n}}A_n(\widehat\btheta_n)$ and $D^2_{h_{2n}}A_{n}(\widehat\btheta_n)$ \\
 Obtain the debiased estimator for $\btheta$ by 
 \[
   \tbtheta_n = \hbtheta_n - \left\{D^2_{h_{2n}}A_{n}(\widehat\btheta_n)\right\}^{-1} D_{h_{1n}}A_n(\widehat\btheta_n)
 \] 
 
Obtain the variance estimator for $\tbtheta_n$ by 
 \[
   \widehat{\mbox{var}}(\tbtheta_n) = n^{-1} \left\{D^2_{h_{2n}}A_{n}(\tbtheta_n)\right\}^{-1} \mathbb{P}_n \left[D_{h_{1n}}m\{\bZ,\cf_n(\tbtheta_n)\}\right]^2 \left\{D^2_{h_{2n}}A_{n}(\tbtheta_n)\right\}^{-1}
 \]
 \caption{Debiased Profile $M$-Estimation (DPME)}
\label{algo:ppme}
\end{algorithm}
%\endgroup

\section{Theoretical Properties}
\label{sec:asymptotic_results}

\subsection{Asymptotic Unbiasedness}

Let $\btheta_0 = \mathfrak{F}(f_0)$ denote the true value of $\btheta$. 
We first assume the following conditions.  
\begin{condition} \label{cond:F_deriv}
The true value $f_{0}$ lies in the interior of $\mathcal{F}$, and $\mathfrak{F}$ has a continuous Hadamard derivative $\nabla\mathfrak{F}(f)[v]$ at $f$ in the neighborhood of $f_{0}$ for any $v \in \mathcal{F}$.
\end{condition}

\begin{condition} \label{cond:m_deriv}
The objective function $m(\bZ, f)$ has a bounded and continuous second-order Hadamard derivative $\nabla^2 m(\bZ, f)[v_1, v_2]$ for $f$ in the neighborhood of $f_{0}$ and any $v_1, v_2\in\mathcal{F}$.
\end{condition}

\begin{condition} \label{cond:f_star}
There exists a unique $f^*(\btheta)$ such that 
$
  f^*(\btheta) = \argmax_{f\in\mathcal{F},\, \mathfrak{F}(f) = \btheta} Pm(\bZ, f).
$
In addition, $f^*(\btheta)$ has a bounded fourth-order derivative with respect to $\btheta$ in the neighborhood of $\btheta_{0}$. 
\end{condition}

\begin{condition} \label{cond:I_0}
The matrix $\bI_0 = - P\nabla^2 m(\bZ, f_0)[\partial_{\btheta} f^*(\btheta_{0}), \partial_{\btheta} f^*(\btheta_{0})]$ is invertible, where $\partial_{\btheta} f^*(\btheta_{0})$ denotes the derivative of $f^*(\btheta)$ with respect to $\btheta$ evaluated at $\btheta_0$.
\end{condition}

Conditions \ref{cond:F_deriv} and \ref{cond:m_deriv} impose smoothness requirements on $\mathfrak{F}(f)$ and $m(\bZ, f)$. 
Condition \ref{cond:f_star} is a standard assumption, similar to those commonly used in profile likelihood \citep{murphy2000profile}. These three assumptions imply that $m\{\bZ,f^*(\btheta)\}$ is fourth-order differentiable with respect to $\btheta$ in the neighborhood of $\btheta_0$, with the first-order derivative $\nabla m\{\bZ,f^*(\btheta)\}[\partial_{\btheta}f^*(\btheta)]$ and second-order derivative $\nabla^2 m\{\bZ,f^*(\btheta)\}[\partial_{\btheta}f^*(\btheta), \partial_{\btheta}f^*(\btheta)]$. Condition \ref{cond:I_0} posits that the expectation of the second-order derivative of $m\{\bZ,f^*(\btheta)\}$ at $\btheta_0$ is non-zero, so that the information matrix $\bI_0$ is positive definite.

The submodel $f^*(\btheta)$ is closely related to the profile likelihood approach \citep{murphy1997maximum, murphy2000profile}. After profiling over nuisance parameters, the relevant derivative behaves like an efficient score, with the ordinary score and information replaced by their efficient counterparts. We illustrate the connection in a simple case where the parameter of interest $\btheta$ and nuisance parameters $\eta$ are separable in $f = (\btheta,\eta)$. The submodel is then $f^*(\btheta) = (\btheta,\eta^*(\btheta))$, where $\eta^*(\btheta)$ is the profiled nuisance parameters that maximizes $Pm\{\bZ,(\btheta,\eta(\btheta))\}$ given $\btheta$. With the submodel $f^*(\btheta)$ replacing the full parameter $(\btheta,\eta)$ in the objective function, the first-order condition on nuisance parameters yields $\partial_{\eta}Pm\{\bZ,(\btheta,\eta^*(\btheta))\}[h]=0$ along any nuisance direction $h$. Hence, under differentiability conditions, the chain rule gives
\begin{align*}
\frac{d}{d\btheta}Pm\{\bZ,f^*(\btheta)\} &= \partial_{\btheta}Pm\{\bZ,(\btheta,\eta^*(\btheta))\} + \partial_{\eta}Pm\{\bZ,(\btheta,\eta^*(\btheta))\}[\partial_{\btheta}\eta^*(\btheta)] \\
&= \partial_{\btheta}Pm\{\bZ,(\btheta,\eta^*(\btheta))\},
\end{align*}
where the left-hand side is the profiled derivative that measures the total contribution of $\btheta$ on the profiled objective function, and the terms in the right-hand side are partial derivatives with respect to the parameter of interest $\btheta$ and nuisance parameters $\eta$. The profiled derivative is Neyman orthogonal to nuisance directions, in the sense that changes in the profiled nuisance do not contribute at first order \citep{foster2023orthogonal}. Therefore, inference on $\btheta$ does not rely on the estimation of nuisance parameters $\eta$ by focusing on the profiled objective function $Pm\{\bZ,f^*(\btheta)\}$.

Since $\nabla\mathfrak{F}(f_0)$ is a bounded linear operator on $\mathcal{F}$, there exists a unique Riesz representor of $\nabla\mathfrak{F}(f_{0})$, which we denote by $\bv^* \in \mathcal{F}^q$, such that $\nabla\mathfrak{F}(f_{0})[v] = \langle \bv^*, v \rangle$ for all $v \in \mathcal{F}$. Here $\langle \bv^*, v \rangle$ represents the componentwise inner product.

\begin{lemma}\label{lemma1}
Under Conditions \ref{cond:F_deriv}--\ref{cond:I_0}, the solution $\bh^* \in \mathcal{F}^q$ to the following equation \begin{equation} \label{equation_h_star}
  P \nabla^2 m(\bZ, f_{0})[\bh^*, v] = \langle \bv^*, v \rangle, \quad\text{ for all } v\in\mathcal{F},
\end{equation}
exists and can be expressed as $\bh^* = - \bI_0^{-1} \partial_{\btheta} f^*(\btheta_{0})$.
\end{lemma}

Note that $\bv^*$ and $\bh^*$ are both vectors of functions with dimensions $q$.
The direction $\bh^*$ plays a role analogous to the least favorable direction in semiparametric efficiency theory \citep{bickel1993efficient}, in that $\nabla m(\bZ, f_0)[\bh^*]$ is ``efficient'', being orthogonal to the tangent space of nuisance parameters. The proof of Lemma~\ref{lemma1} is given in Supplementary Material A.

Now we examine the bias of the initial estimator $\hbtheta_n$.
Suppose that $\widehat{f}_n$ converges to $f_0$ in some distance $d(\cdot, \cdot)$ induced by the inner product $\langle \cdot, \cdot \rangle$. By Taylor expansion,
\[
  \hbtheta_n-\btheta_0 = \mathfrak{F}(\widehat{f}_n)-\mathfrak{F}(f_0) = \langle \bv^*, \widehat{f}_n-f_0 \rangle+o_p\{d(\widehat{f}_n, f_0)\}.
\] 
In addition, since $f_0$ maximizes $P m(\bZ, f)$, we apply another Taylor expansion to obtain
\begin{align*}
  P\nabla  m(\bZ, \widehat{f}_n)[\bh^*] ={} & P\nabla m(\bZ, \widehat{f}_n)[\bh^*] - P\nabla m(\bZ, f_0)[\bh^*] \\
  ={} & P\nabla^2 m(\bZ, f_0)[\bh^*, \widehat{f}_n-f_0]+o_p\{d(\widehat{f}_n, f_0)\} \\
  ={} & \langle \bv^*, \widehat{f}_n-f_0 \rangle+o_p\{d(\widehat{f}_n, f_0)\}.
\end{align*}
Thus, the bias of $\hbtheta_n$ is asymptotically equivalent to $P\nabla m(\bZ, \widehat{f}_n)[\bh^*]$. %which may not be negligible in high-dimensional settings.
In high-dimensional settings, this term may converge at a rate slower than $n^{-1/2}$, making the bias non-negligible.
It follows from Lemma~\ref{lemma1} that 
\begin{align*}
P\nabla m(\bZ, \widehat{f}_n)[\bh^*] = \left\{P\nabla^2 m\{\bZ, f^*(\btheta_{0})\}[\partial_{\btheta} f^*(\btheta_{0}), \partial_{\btheta} f^*(\btheta_{0})]\right\}^{-1}\bigl\{P\nabla m(\bZ, \widehat{f}_n)[\partial_{\btheta} f^*(\btheta_{0})]\bigr\}.
\end{align*}
It is natural to estimate the right-hand side empirically by replacing $P$, $f^*(\btheta)$, and $\btheta_{0}$ with $\mathbb{P}_n$, $\cf_n(\btheta)$, and $\hbtheta_n$, respectively. 
With the derivatives approximated via numerical differentiation, the bias of $\hbtheta_n$ is estimated by $\{D^2_{h_{2n}}A_{n}(\hbtheta_n)\}^{-1} D_{h_{1n}}A_n(\hbtheta_n)$, which is exactly the term subtracted from $\hbtheta_n$ in the one-step update. Provided that the bias estimator is consistent, the debiased estimator $\tbtheta_n$ will be asymptotically unbiased.

\subsection{Asymptotic Normality}

Let $\mathbb{G}_n = n^{1/2}(\mathbb{P}_n - P)$ denote the empirical process.
We further assume the following conditions to establish the asymptotic theory for the proposed debiased estimator $\tbtheta_n$. 

\begin{condition} \label{cond:rate_fhat}
The penalized estimator $\cf_n(\btheta)$ satisfies $d\{\cf_n(\btheta), f^*(\btheta)\} = o_p(n^{-1/4})$ for $\btheta$ in the neighborhood of $\btheta_0$.
\end{condition}

% \begin{condition} \label{cond:d_theta_hat_theta_0}
% The initial estimator $\htheta_n$ satisfies $\|\htheta_n - \btheta_{0}\| = o_p(n^{-1/4})$.
% \end{condition}

\begin{condition} \label{cond:asymp_equicont}
For every $\epsilon, \eta > 0$, there exist $\delta_1, \delta_2 > 0$ such that
  $$\lim_{n\to\infty} \pr \left( \sup_{f_1,f_2 \in \mathcal{N}_{f_{0},\delta_1}; \bh_1,\bh_2 \in \mathcal{N}_{\bh^*, \delta_2}} \left\| \mathbb{G}_n \{ \nabla m(\bZ, f_1)[\bh_1]\} - \mathbb{G}_n \{ \nabla m(\bZ, f_{2})[\bh_2] \} \right\| > \epsilon \right) < \eta,$$ where $\mathcal{N}_{f_0, \delta} = \{ f : d(f , f_0) \leq \delta \}$ and $\mathcal{N}_{\bh^*, \delta} = \{ \bh : \|d(\bh , \bh^*)\| \leq \delta \}$.
\end{condition}

\begin{condition} \label{cond:consistent}
For every $\eta > 0$, there exist $\epsilon, \delta_1, \delta_2 > 0$ such that
$$\lim_{n\to\infty} \pr \left(\sup_{f \in \mathcal{N}_{f_{0},\delta_1}; \bh_1,\bh_2 \in \mathcal{N}_{\bh^*, \delta_2}} n^{1/4} \|\mathbb{P}_n\nabla^2 m(\bZ,f)[\bh_1,\bh_2] - P\nabla^2 m(\bZ,f)[\bh_1,\bh_2]\| > \epsilon\right) < \eta,$$ where $\mathcal{N}_{f_0, \delta} = \{ f : d(f , f_0) \leq \delta \}$ and $\mathcal{N}_{\bh^*, \delta} = \{ \bh : \|d(\bh , \bh^*)\| \leq \delta \}$.
\end{condition}

\begin{condition} \label{cond:mdonsker}
For some constant $c$, the function classes $\mathcal{C}_1 = \{\partial_{\btheta}m(\bZ,f^*(\btheta)): \|\btheta-\btheta_0\|<c\}$ and $\mathcal{C}_2 = \{\partial^2_{\btheta\btheta}m(\bZ,f^*(\btheta)): \|\btheta-\btheta_0\|<c\}$ are Donsker.
\end{condition}

\begin{condition} \label{cond:num_deriv}
$\|D_{h_{1n}}m\{\bZ,f^*(\btheta)\} - \partial_{\btheta}m\{\bZ,f^*(\btheta)\}\|_{L_2(P)} \to_p 0$ and $\|D^2_{h_{2n}}m\{\bZ,\cf_n(\btheta)\} - \partial^2_{\btheta\btheta}m\{\bZ,f^*(\btheta)\}\|_{L_2(P)} \to_p 0$ in the neighborhood of $\btheta_0$.
\end{condition}

Condition \ref{cond:rate_fhat} requires that the convergence rate of $\cf_n$ be faster than $n^{-1/4}$, which is typically achievable under specific sparsity or smoothness conditions.
Combined with the smoothness of $\mathfrak{F}(f)$ specified in Condition \ref{cond:F_deriv}, this condition ensures that $\|\htheta_n - \btheta_{0}\| = o_p(n^{-1/4})$. For example, in lasso regression with sparsity $s_0$, the error of $\cf_n=\cf_n(\hbtheta_n)$ compared to $f_0=f^*(\btheta_0)$ is of the order $O_p(s_0^{1/2}\lambda_n)$. With the typical tuning parameter $\lambda_n = (\log p/n)^{1/2}$, Condition \ref{cond:rate_fhat} holds if $s\log p = o(n^{1/2})$.
Condition \ref{cond:asymp_equicont} requires the asymptotic equicontinuity of the empirical process $\mathbb{G}_n \{ \nabla m(\bZ, f)[\bh]\}$ in neighborhoods of $f_0$ and $\bh^*$. 
Condition \ref{cond:consistent} requires the empirical second-order derivative of the $m$-function to be consistent at a rate of $O_p(n^{-1/4})$ in the neighborhoods of $f_0$ and $\bh^*$, which implies $(\mathbb{P}_n-P)\partial^2_{\btheta\btheta} m\{\bZ,f^*(\hbtheta_n)\} = O_p(n^{-1/4})$. This condition is weaker than the condition underlying the law of large numbers. Condition \ref{cond:mdonsker} requires that the derivatives of the profile $m$-function are not too complex. While the Donsker condition may fail for the function class indexed by $f\in\mathcal{F}$, the profiled function class indexed by $\{\btheta: \|\btheta-\btheta_0\|<c\}$ is much simpler and can satisfy Donskerity more easily, as the derivatives only concern the neighborhood of the least favorable direction. Condition \ref{cond:num_deriv} states that the first- and second-order numerical derivatives based on $\cf_n(\btheta)$ should approximate the true derivatives of the $m$-function with respect to $\btheta$ around $\btheta_0$. 

For debiasing, a good approximation for the first-order derivative is important; the reason behind this is provided in Supplementary Material B. In particular, the error of the approximated first-order derivative should be $o_p(n^{-1/2})$, while the error of the second-order derivative can be $O_p(n^{-1/4})$. Since the first-order derivative is calculated by the centralized numerical differentiation as shown in Equation \eqref{D1_cal}, a truncation error of $O_p(h_{1n}^2)$ is incurred. Therefore, $h_{1n}$ should be $o_p(n^{-1/4})$. Another consideration for choosing $h_{1n}$ is the smoothness of the first-order numerical derivative. Due to penalization, $\cf_n(\btheta)$ may not be differentiable with respect to $\btheta$ at $\hbtheta_n$, which is the case of lasso. To avoid unstable numerical derivatives, $h_{1n}$ should be large enough so that the error of $d\{\cf_n(\hbtheta_n \pm h_{1n}),f^*(\hbtheta_n \pm h_{1n})\}$ is negligible compared to $h_{1n}$. This can be guaranteed by a properly chosen $h_{1n}$ that satisfies $d(\cf_n,f_0) = o_p(h_{1n})$. As shown in Supplementary Material B, for $h_{1n}=o_p(n^{-1/4})$, $d(\cf_n,f_0) = o_p(h_{1n})$, $h_{2n}=O_p(n^{-1/4})$, and $d(\cf_n,f_0) = o_p(h_{2n})$,
\begin{align*}
D_{h_{1n}}A(\hbtheta_n) &= \mathbb{P}_n \partial_{\btheta} m\{\bZ,f^*(\hbtheta_n)\} + o_p(n^{-1/2}), \\
D^2_{h_{2n}}A(\hbtheta_n) &= P \partial^2_{\btheta\btheta} m\{\bZ,f^*(\hbtheta_n)\} + O_p(n^{-1/4}).
\end{align*}
In practice, $h_{jn}$ can be chosen at $c_j n^{-1/4-\varepsilon}$ for constants $c_j$ and a constant $\varepsilon$ that is close to zero ($j=1,2$). If $\btheta$ is a scalar, we can choose $h_{1n}=2h_{2n}=c_1n^{-1/4-\varepsilon}$ so the penalized regression is additionally performed at only two points, $\hbtheta_n \pm h_{1n}$. When iterative algorithms are used for optimization, the solver tolerance must be $o_p(n^{-1/2}h_{1n}+n^{-1/2}h_{2n})$ for derivatives. 
%If $\cf_n(\btheta)$ is differentiable with respect to $\btheta$, the error of approximating $\mathbb{P}_n\partial_{\btheta}m\{\bZ,f^*(\hbtheta_n)\}$ using $\cf_n(\hbtheta_n)$ is $O_p(\|\partial_{\btheta}\cf_n(\htheta_n)-\partial_{\btheta}f^*(\htheta_n)\|^2)$. In this case, Conditions \ref{cond:mdonsker} and \ref{cond:num_deriv} can be replaced by $\|\partial_{\btheta}\cf_n(\htheta_n)-\partial_{\btheta}f^*(\htheta_n)\| = o_p(n^{-1/4})$.

The following theorem establishes that the debiased estimator $\widetilde\btheta_n$ is regular and asymptotically linear. 

\begin{theorem}\label{thm1}
    Under Conditions \ref{cond:F_deriv}--\ref{cond:num_deriv}, $$\sqrt{n}(\tbtheta_n - \btheta_{0}) = -\mathbb{G}_n \left\{ \nabla m(\bZ, f_{0})[\bh^*] \right\} + o_p(1).$$
\end{theorem}

The proof of Theorem~\ref{thm1} is given in Supplementary Material B.
By Lemma~\ref{lemma1}, we have 
\[
\mbox{var}\left\{\nabla m(\bZ, f_{0})[\bh^*]\right\} = \bm{I}_0^{-1} \mbox{var}\left\{\nabla m(\bZ, f_{0})[\partial_{\btheta} f^{*}(\btheta_{0})]\right\} \bm{I}_0^{-1}.
\]
Since $\nabla m(\bZ, f_{0})[\partial_{\btheta} f^{*}(\btheta_{0})]$ has mean zero, 
we can estimate its variance-covariance matrix by $\mathbb{P}_n [D_{h_{1n}}m\{\bZ,\cf_n(\tbtheta_n)\}]^2$.
In addition, we estimate $\bm{I}_0$ by $-D^2_{h_{2n}}A_{n}(\tbtheta_n)$. 
These estimators yield the sandwich variance estimator for $\tbtheta_n$ given in \eqref{variance_estimator} in Section~\ref{sec:Method}.
Under the conditions of Theorem \ref{thm1}, this variance estimator is consistent for the asymptotic variance of $\tbtheta_n$, even if we do not know the explicit form of $\bh^*$.

Interestingly, our asymptotic theory does not require any assumptions on penalization, making it broadly applicable to a wide range of estimation problems. 
As long as the initial estimator $\cf_n$ converges to the true value $f_0$ at a rate faster than $n^{-1/4}$, the proposed DPME approach can be used to obtain a regular and asymptotically linear estimator for $\btheta$.

\section{Case Studies}\label{examples}

\subsection{High-Dimensional Linear Models (Lasso)}

Suppose that the sample includes $n$ i.i.d. copies of $\bZ = (\bX, Y)$, where $\bX \in \mathbb{R}^{p_n}$ is the covariates and $Y$ is the response variable. The response variable is modeled by a linear model $Y = \bX^{\top} \bbeta + \epsilon$, where $\epsilon$ is a random error. The complete functional parameter $f = \bbeta$ includes all arguments of $\bbeta$. For illustration purposes, suppose that we are interested in the first argument of $\bbeta$, denoted by $\beta_1 = \mathfrak{F}(\bbeta)$.
The true parameter $\bbeta_0$ maximizes $Pm(\bZ,\bbeta)$, where $m(\bZ, \bbeta) = -(Y - \bX^\top \bbeta)^2/2$. We denote $\bbeta_0 = (\beta_{1,0}, \bbeta_{-1,0}^{\top})^{\top}$. 

We first obtain an initial estimator by Lasso,
\begin{equation*}
    \widehat{\bbeta}_{n} = \argmax_{\bbeta}   -\frac{1}{2} \mathbb{P}_n \left\{ (Y - \bX^\top \bbeta)^2 \right\}  - \lambda_n \|\bbeta\|_1.
\end{equation*}
The tuning parameter $\lambda_n$ is selected through cross-validation. The initial estimator for $\beta_1$, denoted by $\widehat\beta_{1,n}$, is the first element of $\widehat\bbeta_n$. Next, under the constraint of $\beta_1 \in \{\widehat\beta_{1,n}\pm h_{1n}, \widehat\beta_{1,n}\pm h_{2n}\}$, we estimate $\bbeta_{-1}$ again by Lasso,
\begin{equation*}
    \widehat{\bbeta}_{-1,n}(\beta_1) = \argmax_{\bbeta_{-1}}   -\frac{1}{2} \mathbb{P}_n \left\{ (Y - X_1 \beta_1 - \bX_{-1}^\top \bbeta_{-1})^2 \right\}  - \lambda_n \|\bbeta_{-1}\|_1,
\end{equation*}
where $\bbeta_{-1}$ is the vector of $\bbeta$ by removing the first element, and $\bX_{-1}$ is the vector of covariates by removing the first element. 
Based on \eqref{D1_cal} and \eqref{D2_cal}, we calculate the first- and second-order numerical derivatives of the profile $m$-function, $D_{h_{1n}}A_n(\widehat\beta_{1,n})$ and $D^2_{h_{2n}}A_n(\widehat\beta_{1,n})$.
Then, the one-step updated estimator is obtained by
\[
\widetilde\beta_{1,n} = \widehat\beta_{1,n} - \{D^2_{h_{2n}}A_n(\widehat\beta_{1,n})\}^{-1} D_{h_{1n}}A_n(\widehat\beta_{1,n}).
\]

We now verify conditions in Theorem \ref{thm1}. The extraction operator $\beta_1 = \mathfrak{F}(\bbeta)$ is a coordinate selector and thus satisfies Condition \ref{cond:F_deriv}. The $m$-function $m(\bZ,\bbeta)$ is second-order differentiable with respect to $\bbeta$, so Condition \ref{cond:m_deriv} holds. The submodel in this case is $f^*(\beta_1) = (\beta_1, \bbeta_{-1}^*(\beta_1)^{\top})^{\top}$, with
\begin{align*}
    \bbeta_{-1}^*(\beta_1) &= \argmax_{\bbeta_{-1}}  -\frac{1}{2} P\left\{ (Y - X_1 \beta_1 - \bX_{-1}^\top \bbeta_{-1})^2 \right\} = \{P(\bX_{-1}\bX_{-1}^{\top})\}^{-1} P\{\bX_{-1} (Y - X_1\beta_{1})\}.
\end{align*}
Provided that $P(\bX_{-1}\bX_{-1}^{\top})$ is invertible, $f^*(\beta_1)$ exists and is unique with continuous derivatives of any order with respect to $\beta_1$, Condition \ref{cond:f_star} is satisfied.
Let $\bm\gamma_0$ be the regression coefficient of $\bX_{-1}$ by regressing $X_1$ on $\bX_{-1}$. Then
$$\partial_{\beta_{1}} \bbeta_{-1}^*(\beta_{1,0}) = -\{P(\bX_{-1}\bX_{-1}^\top)\}^{-1} P\{X_1 \bX_{-1}\} = (1, -\bm\gamma_0^\top )^\top.$$
By some simple calculation, we have
$$\partial^2_{\beta_1\beta_1} m(\bZ,f^*(\beta_{1,0})) = -(1, -\bm\gamma_0^\top ) \bX \bX^{\top}  (1, -\bm\gamma_0^\top )^\top$$
and 
$I_0 = -P \partial^2_{\beta_1\beta_1} m(\bZ,f^*(\beta_{1,0})) = (1, -\bm\gamma_0^\top ) P(\bX\bX^{\top}) (1, -\bm\gamma_0^\top )^\top.$
Condition \ref{cond:I_0} is satisfied if $X_1$ is not a linear function of $\bX_{-1}$, so that $I_0 \neq 0$ is invertible. Let $h^* = -I_0^{-1} (1,-\bm\gamma_0^{\top})^{\top}$ and it follows that
\[
\nabla m(\bZ, \bbeta_0)[h^*] = - I_0^{-1} \bX^{\top} (1, -\gamma_0^\top )^\top (Y- \bX^\top \bbeta_0).
\]
In Lasso, the direction $h^*$ has an explicit form. The debiasing Lasso estimates the $\gamma_0$ in $h^*$ using a Lasso regression for $X_1$ to $\bX_{-1}$. In contrast, DPME numerically solves this direction by perturbing $\beta_1$.

In the initial Lasso, with high probability, $\| \widehat{\bbeta}_{n} - \bbeta_{0}\|^2_2 \leq C s_0 n^{-1} \log p_n$ for some constant $C$, where $s_0 = \|\bbeta_0\|_0$ is the sparsity level \citep{van2014asymptotically}. If $s_0 = o(\sqrt{n}/\log p_n)$, then $\| \widehat{\bbeta}_{n} - \bbeta_0 \|_2 = o_p(n^{-1/4})$, which implies Condition \ref{cond:rate_fhat}. Using the maximal inequality, $P(\|\mathbb{G}_n\{\nabla m(\bZ,\bbeta)[h]\}\|>x)$ can be shown to be bounded by the entropy $\int_{0}^{\delta_n} \sqrt{\log N(\varepsilon,H_n,d_n)} d\varepsilon$ for some $\delta_n$ and some space $H_n$ up to a constant \citep{wang2024one}. Under the conditions in \citet{bartlett2012regularized}, this entropy goes to zero, so Condition \ref{cond:asymp_equicont} holds.
Suppose $(1,-\bm\gamma_0^{\top})\bX$ has a bounded variance, then $\mathbb{P}_n\partial^2_{\beta_1\beta_1}m(\bZ,f^*(\beta_1)) = -I_0 + O_p(n^{-1/2})$, implying Condition \ref{cond:consistent}. Suppose that $P\|\bX\|^4<\infty$, $P(Y^4)<\infty$, $\|\bm\gamma_0\|<\infty$, then $\partial_{\beta_1}m(\bZ,f^*(\beta_1))$ and $\partial^2_{\beta_1\beta_1}m(\bZ,f^*(\beta_1))$ are Donsker, implying Condition \ref{cond:mdonsker}. Choosing $h_{1n}=2h_{2n}=c_1n^{-1/4-\epsilon}$ for a constant $c_1$ and a small enough $\epsilon$ ensures Condition \ref{cond:num_deriv}.

In particular, if the linear model is correctly specified and $\epsilon \sim N(0,\sigma^2)$, then
\[
\sqrt{n} (\widetilde\beta_{1,n} - \beta_{1,0}) = -\mathbb{G}_n \{\nabla m(\bZ, f_{0})[h^*]\} + o_p(1) \rightarrow_{d} N(0, \sigma^2 I_0^{-1}).
\]
The variance of $\widetilde\beta_{1,n}$ is numerically estimated using the derivatives at $\widetilde\beta_{1,n}$ by perturbing $\widetilde\beta_{1,n}$ based on \eqref{variance_estimator}.

%Suppose $s_\Omega$ is the maximum sparsity level of the rows of $\Omega = \Sigma^{-1}$ and $s_\Omega = O(n^{\alpha_1}/ \log p_n)$, where $\alpha_1<1/2$. This implies that $\gamma_0$ is sparse and this is Condition B.5 of Wang et al. 2024. According to the proof of Wang et al. 2024, we know that we only need to prove$$\mathbb{G}_n \{ \nabla m(\bZ, f_0)[h_0]\} = -(Y- \bX^\top \bbeta)I_0^{-1} \bX (1, -\gamma_0^\top )^\top. $$

\subsection{High-Dimensional Logistic Models for Binary Outcomes}

Suppose that the sample includes $n$ i.i.d. copies of $\bZ = (\bX, Y)$ with $\bX \in \mathbb{R}^{p_n}$ and $Y_i \in \{0, 1\}$. We assume that the last element of $\bX$ is 1, and all other elements have a mean of zero. We assume that $Y$ follows a Bernoulli distribution with mean $P(Y = 1 \mid \bX) = G\left(\bX^{\top}\bbeta\right)$, where $G$ is the link function with a continuous third-order derivative, and $\bbeta \in \mathbb{R}^{p_n}$ is the vector of unknown parameters. Specifically, $G(x) = \exp(x)/\{1+\exp(x)\}$ for logistic regression.
We define the $m$-function as the log-likelihood function,
\[
m(\bZ, \bbeta) = Y \log \left\{ \frac{ G\left(\bX^{\top}\bbeta\right)}{ 1 - G\left(\bX^{\top}\bbeta\right)} \right\} + \log\left\{ 1 -  G\left(\bX^{\top}\bbeta\right)\right\}.
\]
The true parameter $\bbeta_0$ maximizes $P m(\bZ,\bbeta)$. Suppose we are interested in the first argument of $\bbeta = (\beta_1,\bbeta_{-1}^{\top})^{\top}$. 

First, we obtain an initial estimator for $\bbeta$ by including a Lasso penalty to the log-likelihood,
\[
\widehat\bbeta_n = \argmax_{\bbeta} \mathbb{P}_n \{m(\bZ,\bbeta)\} - \lambda_n \|\bbeta\|_1.
\]
The tuning parameter $\lambda_n$ is selected through cross-validation. 
Next, given the initial estimate $\widehat\beta_{1,n}$, we fit constrained logistic-Lasso at $\beta_1 \in \{\widehat\beta_{1,n}\pm h_{1n}, \widehat\beta_{1,n}\pm h_{2n}\}$ to obtain $\widehat\bbeta_n(\beta_1) = (\beta_1,\widehat\bbeta_{-1,n}(\beta_1)^{\top})^{\top}$ by
\[
\widehat\bbeta_{-1,n}(\beta_1) = \argmax_{\bbeta_{-1}} \mathbb{P}_n \{m(\bZ,(\beta_1,\bbeta_{-1}^{\top})^{\top})\} - \lambda_n \|\bbeta_{-1}\|_1.
\]
We compute the first- and second-order numerical derivatives $D_{h_{1n}}A_n(\widehat\beta_{1,n})$ and $D^2_{h_{2n}}A_n(\widehat\beta_{1,n})$ based on \eqref{D1_cal} and \eqref{D2_cal}. Then, the one-step updated estimator is constructed by
\[
\widetilde\beta_{1,n} = \widehat\beta_{1,n} - \{D^2_{h_{2n}}A_n(\widehat\beta_{1,n})\}^{-1} D_{h_{1n}}A_n(\widehat\beta_{1,n}).
\]
The DPME estimate has a similar form to existing debiasing estimators \citep{van2014asymptotically}. However, the computation of DPME is much more efficient because only numerical derivatives are required for debiasing, avoiding solving the explicit form of the debiasing direction $h^*$.

Now we verify that the conditions in Theorem \ref{thm1} are satisfied. Similar to the linear model, Conditions \ref{cond:F_deriv} and \ref{cond:m_deriv} hold. The submodel in this case is $f^*(\beta_1) = (\beta_1, \bbeta_{-1}^*(\beta_1))$, with
\begin{equation*}
	\bbeta_{-1}^*(\beta_1) = \argmax_{\bbeta_{-1}} P \left\{ Y \log \left[ \frac{ G\left(X_{1} \beta_{1} + \bX_{-1}^{\top} \bbeta_{-1}\right)}{ 1 - G\left(X_{1} \beta_{1} + \bX_{-1}^{\top} \bbeta_{-1}\right)} \right] + \log\left[ 1 -  G\left(X_{1} \beta_{1} + \bX_{-1}^{\top} \bbeta_{-1}\right)\right] \right\}.
\end{equation*}
Since the objective function is differentiable, by taking the derivative with respect to $\bbeta_{-1}$, we know that $\bbeta_{-1}^*(\beta_1)$ satisfies
\begin{equation} \label{eq:glm}
    P \left\{ l'(Y, X_1\beta_{1} + \bX_{-1}^{\top}\bbeta^*_{-1}(\beta_1) ) \bX_{-1}\right\} = 0,
\end{equation}
where
$$l'(y, x) = y \frac{G'(x)}{G(x)\left(1 - G(x)\right)} - \frac{G'(x)}{1 - G(x)}.$$
For the logistic model, the solution to the equation above exists and is unique, so Condition \ref{cond:f_star} holds.
By taking the derivative with respect to $\beta_1$ in Equation \eqref{eq:glm},
\begin{equation} \label{eq:glm2}
\begin{aligned}
	&P \left\{ l''\left(Y, X_1\beta_1+\bX_{-1}^{\top}\bbeta_{-1}^*(\beta_1) \right)X_{1} \bX_{-1}\right\} \\
    & + P\left\{ l''\left(Y, X_1\beta_1+\bX_{-1}^{\top}\bbeta_{-1}^*(\beta_1) \right)\bX_{-1} \bX^{\top}_{-1}\right\} \partial_{\beta_{1}}\bbeta_{-1}^*(\beta_{1}) = 0,
\end{aligned}
\end{equation}
where $l''(y, x) = yH_1(x) - H_2(x)$ with
\begin{align*}
H_1(x) &= \frac{G''(x)G(x)\left(1 - G(x)\right) - \left(1 - 2G(x)\right)G'(x)^2}{G(x)^2\left(1 - G(x)\right)^2}, \\
H_2(x) &= \frac{G''(x)(1 - G(x)) + G'(x)^2}{\left(1 - G(x)\right)^2}.
\end{align*}
Therefore, $\partial_{\beta_1}\bbeta_{-1}^*(\beta_1)$ exists. By taking the derivative again, we can verify that Condition \ref{cond:I_0} is satisfied if $P\{l''\left(Y, \bX^{\top}\bbeta \right)(\bX_{-1} \bX^{\top}_{-1})\}$ is invertible. However, the explicit form of $I_0$ is complex. Since $l'(Y,\bX^{\top}\bbeta)$ is Lipschitz continuous in $\bbeta$, Condition \ref{cond:asymp_equicont} holds for bounded $\bX$ under some regularity conditions \citep{bartlett2012regularized}. Under the finite second-order moment condition of $\bX$, we have $(\mathbb{P}_n-P)\partial^2_{\beta_1\beta_1}m\{\bZ,f^*(\beta_1)\}\ = O_p(n^{-1/2})$, which implies Condition \ref{cond:consistent}.

In the initial logistic-Lasso, with high probability, $\|\widehat\bbeta_n - \bbeta_0\|_2^2 \leq C s_0 n^{-1} \log p_n$ for some constant $C$, where $s_0$ is the sparsity level \citep{blazere2014oracle}. If $s_0 = o(\sqrt{n}/\log p_n)$, then Condition \ref{cond:rate_fhat} is satisfied. Let $\bm\gamma_0$ be the regression coefficient for $X_1$ on $\bX_{-1}$ with weight $l''(Y,\bX^{\top}\bbeta)$. Suppose that $P\|\bX\|^2<\infty$, $\|\bm\gamma_0\|<\infty$, then $\partial_{\beta_1}m(\bZ,f^*(\beta_1))$ and $\partial^2_{\beta_1\beta_1}m(\bZ,f^*(\beta_1))$ are Donsker, implying Condition \ref{cond:mdonsker}. Choosing $h_{1n}=2h_{2n}=c_1n^{-1/4-\epsilon}$ for a constant $c_1$ and a small enough $\epsilon$ ensures Condition \ref{cond:num_deriv}.

\subsection{High-Dimensional Individualized Treatment Rules} \label{case_itr}

In the previous two examples, we need to check the Donskerity of the profile $m$-function derivatives in a case-by-case manner. When there are estimated nuisance parameters in the model, Donsker conditions are difficult to verify or even fail. The modern double/debiased machine learning avoids Donsker conditions through sample splitting and cross-fitting \citep{chernozhukov2018double}. The complete sample is divided into two parts, one for estimating nuisance parameters and the other for making inference. In this section, we consider estimating individualized treatment rules (ITRs), which involves estimating nuisance parameters.

Let $\bX \in \mathbb{R}^{p_n}$ be the covariates, $A \in \{-1, 1\}$ be the treatment assignment, and $Y(a)$ be the potential outcome, $a \in \{-1, 1\}$, where we assume higher values of $Y(a)$ indicate better clinically status. 
An individualized treatment rule, denoted by $D(\bX)$, is a mapping from the space of covariates to the space of treatments. Under consistency, the observed outcome under the decision rule $D(\bX)$ is $Y(D) = \sum_{a\in\{-1,1\}} Y(a)I\{a = D(\bX)\}$. The value function $V(D) = E(Y(D))$ is the mean outcome in the population that would be obtained if the decision rule $D(\bX)$ were implemented. The optimal ITR is defined as $D_{\text{opt}} = \arg \max_D \{V(D)\}$.
We assume the stable unit treatment value, strong ignorability, and positivity. Under these assumptions, we can identify and estimate the value function from i.i.d. observations of $\bZ = (A,Y,\bX)$ \citep{zhao2012estimating}.

We model the optimal ITR by $D(\bX) = \mathrm{sign}(\bX^{\top}\bbeta)$ \citep{xu2015regularized}. The complete parameter is $f = \bbeta$, and we may be interested in some (or all) elements of $\bbeta$. To remove the dependency between estimating ITR and nuisance models, we adopt the sample splitting method proposed in \cite{liang2022estimation}. We randomly divide the sample into a training set and an inference set, with each set approximately half the sample size. In the training set, a distance correlation-based variable screening procedure is applied to fit the propensity score $\pi(\bX; a) = P(A=a \mid \bX)$ and the outcome model $\mu(\bX; a) = E(Y \mid \bX, A=a)$ using kernels \citep{li2012feature}, denoted by $\widehat\pi(\bX; a)$ and $\widehat\mu(\bX; a)$, respectively. In the inference set, we compute the pseudo-outcome $\widehat{Y}(a)$ by augmented inverse probability weighting 
\[
\widehat{Y}(a) = \frac{I(A=a)}{\widehat\pi(\bX; a)}\{Y - \widehat\mu(\bX; a)\} + \widehat\mu(\bX; a).
\]
We denote $\mathbb{P}_n^{(2)}$ as the sample average in the inference set. The value function $V(D)$ can be consistently estimated by $\widehat{V}(D) = \mathbb{P}_n^{(2)}[\widehat{Y}(1)I\{D(\bX)=1\} + \widehat{Y}(-1)I\{D(\bX)=-1\}]$ if either the propensity score or the outcome model is correctly specified. 

By considering the positive and negative parts separately, maximizing $\widehat{V}(D)$ is equivalent to minimizing 
\[
\mathbb{P}_n^{(2)} [\{\widehat{Y}(1)_{+}+\widehat{Y}(-1)_{-}\} I\{D(\bX)\neq1\} + \{\widehat{Y}(1)_{-}+\widehat{Y}(-1)_{+}\} I\{D(\bX)\neq-1\}].
\]
We define weights $\widehat\Omega_{+} = \widehat{Y}(1)_{+}+\widehat{Y}(-1)_{-}$ and $\widehat\Omega_{-} = \widehat{Y}(1)_{-}+\widehat{Y}(-1)_{+}$. A large $\widehat\Omega_{+}$ and a small $\widehat\Omega_{-}$ suggest that the treatment is beneficial. By replacing the 0-1 loss with logistic loss $\phi(t) = \log(1+e^{-t})$, an initial estimator for $\bbeta$ is obtained by 
\begin{align*}
\widehat\bbeta_n^{(2)} &= \argmin_{\bbeta} \mathbb{P}_n^{(2)} [\widehat\Omega_{+} \phi(\bX^{\top}\bbeta) + \widehat\Omega_{-} \phi(-\bX^{\top}\bbeta)] - 2\lambda_n\|\bbeta\|_1 \\
&= \argmax_{\bbeta} \mathbb{P}_n^{(2)} [\widehat\Omega_{+}\bX^{\top}\bbeta - \widehat\Omega_{+} \phi(-\bX^{\top}\bbeta) - \widehat\Omega_{-} \phi(-\bX^{\top}\bbeta)] - 2\lambda_n\|\bbeta\|_1.
\end{align*}
We note that it has the same loss function as the logistic-Lasso. For each individual, we pretend there is a replica. We use $(1, 0)$ as the response variable, $(\bX, \bX)$ as the covariates, and $(\widehat\Omega_{+}, \widehat\Omega_{-})$ as the weight to fit logistic-Lasso. The tuning parameter is selected through cross-validation.

To debias $\widehat\bbeta_n^{(2)}$, we perturb each argument of $\widehat\bbeta_n^{(2)}$ at one time, say, $\beta_{j} \in \{\widehat\beta_{j,n}^{(2)} \pm h_{1n}, \widehat\beta_{j,n}^{(2)} \pm h_{2n}\}$, and compute $\widehat\bbeta_{-j,n}^{(2)}(\beta_j)$ by logistic-Lasso,
\[
\widehat\bbeta_{-j,n}^{(2)}(\beta_j) = \argmax_{\bbeta_{-j}} -\mathbb{P}_n^{(2)}[\widehat\Omega_{+}\phi(\bm{X}_{-j}^{\top}\bbeta_{-j}+X_j\beta_j) + \widehat\Omega_{-}\phi(-\bX_{-j}^{\top}\bbeta_{-j}-X_j\beta_j)] - 2\lambda_n\|\bbeta_{-j}\|_1.
\]
Then we compute the numerical derivatives $D_{h_{1n}}A_n(\widehat\bbeta_{j,n}^{(2)})$ and $D^2_{h_{2n}}A_n(\widehat\bbeta_{j,n}^{(2)})$. The debiased estimator for $\beta_j$ is obtained by 
\[
\widetilde\beta_{j,n}^{(2)} = \widehat\beta_{j,n}^{(2)} - D^2_{h_{2n}}A_n(\widehat\bbeta_{j,n}^{(2)})^{-1} D_{h_{1n}}A_n(\widehat\bbeta_{j,n}^{(2)}).
\]
To improve estimation efficiency, cross-fitting is applied by switching training and inference sets, which gives another debiased estimator $\widetilde\bbeta_n^{(1)}$. The final estimator for $\bbeta$ is given by $\widetilde\bbeta_n = (\widetilde\bbeta_n^{(1)} + \widetilde\bbeta_n^{(2)})/2$.

To justify this procedure, let $\eta$ represent nuisance parameters in the propensity score and outcome model, whose true value is $\eta_0$. Let $\Omega_{+} = \Omega_{+}(\eta_0)$ and $\Omega_{-} = \Omega_{-}(\eta_0)$ be the weights by plugging in the true propensity score and outcome model. 
The $m$-function $m(\bZ,f,\eta) = \Omega_{+}(\eta)\phi(\bX^{\top}\bbeta) + \Omega_{-}(\eta)\phi(-\bX^{\top}\bbeta)$. We define $\bbeta_0$ as the vector such that
\[
\bbeta_0 = \argmin_{\bbeta} P[\Omega_{+}(\eta_0)\phi(\bX^{\top}\bbeta) + \Omega_{-}(\eta_0)\phi(-\bX^{\top}\bbeta)].
\]
For identifiability, we restrict the $k$th element $|\beta_{k,0}|=1$, assuming $X_k$ is a significant covariate in the ITR. The convergence rates of $\widehat\Omega_{+} = \Omega_{+}(\widehat\eta_n)$ and $\widehat\Omega_{-} = \Omega_{-}(\widehat\eta_n)$ can achieve $o_p(n^{-1/4})$ after variable screening. 
Due to sample splitting, we can treat the weights $(\widehat\Omega_{+}, \widehat\Omega_{-})$ as fixed when debiasing $\widehat\beta_{n}^{(2)}$ in the inference set. Under some bound conditions on $\|\bbeta_0\|_1$, the interaction between $\widehat\eta_n-\eta_0$ and $\bX^{\top}(\widehat\bbeta_n^{(2)}-\bbeta_0)$ can be bounded by $O_p(s_0\log(np_n)n^{-3/4})$ and is negligible compared to $\widehat\bbeta_n^{(2)}-\bbeta_0$. Hence, the error in the estimated weights has no first-order influence on $\widehat\bbeta_n^{(2)}$. Therefore, Conditions \ref{cond:asymp_equicont}, \ref{cond:consistent}, \ref{cond:mdonsker}, and \ref{cond:num_deriv} for establishing asymptotic properties for the debiased estimator pertain to the case of weighted logistic models with fixed weights, which are straightforward extensions of the previous section. Under the regularity conditions in \citet{liang2022estimation}, $\|\widehat\bbeta_n^{(2)} - \bbeta_0\|_1 = O_p(s_0\sqrt{\log p_n/n})$. Provided that $s_0$ is not too large, we have $\|\widehat\bbeta_n^{(2)} - \bbeta_0\| = o_p(n^{-1/4})$, ensuring Condition \ref{cond:rate_fhat}.

\section{Simulation Studies}\label{simulation}

\subsection{Simulation 1: High-Dimensional Linear Models} \label{sec_lasso}

The first simulation study examines a high-dimensional linear model. In this setup, we generate $p \in \{100, 500\}$ covariates $\bX$, which are divided into $K = p/4$ groups, with each group containing $4$ variables. For the variables in the $k$th group, denoted as $X_{k1},\ldots, X_{k4}$, we generate them as follows:
$$X_{kj}=0.5(w_{kj} + tu_{k}), \quad w_{kj}\sim U(0,1), \ u_k\in U(0,1).$$
This approach allows us to generate a sequence of blocked covariates, with a correlation of $\rho = 0.5$ between any two $X$'s within the same block, while they remain independent if they belong to different blocks. Given $\bX$, the outcome is generated from a linear model $Y = \bX^{\top} \boldsymbol{\beta}_0 + \epsilon$ with $\bbeta_0 = (1,1,1,1,1,0,\dots,0)^{\top}$ and $\epsilon \sim N(0,1)$.

To demonstrate the effectiveness of our proposed method, as an illustration, we consider inference for the coefficient of a significant covariate $\beta_1$ and the coefficient of a non-significant covariate $\beta_6$. We investigate sample sizes of $n=200, 500, 1000$, with each scenario replicated 500 times. The simulation is conducted on a laptop with a 13th Gen Intel(R) Core(TM) i7-1365U 1.80GHz processor.
We initially estimate the values of $\bbeta$ through Lasso regression using the ``glmnet'' package in R. The tuning parameter is selected via cross-validation as the $\lambda_n$ that yields the absolute lowest cross-validation error.
We choose $h_{1n}=2h_{2n}=0.75n^{-0.26}$, so that we only need to perform constrained penalized regressions at $\widehat\beta_{1,n} \pm h_{1n}$. The tuning parameter $\lambda_n$ is held unchanged in the constrained penalized regressions. Additionally, we compare our method with debiased Lasso in the ``desla'' package \citep{van2014asymptotically}.
%Simulation found that a smaller $h_1n$ leads to smaller finite-sample variation but larger finite-sample bias; a larger $h_{2n}$ is preferred.

Table \ref{Table_Lasso} presents the simulation results for the coefficients of one important predictor ($\beta_1$) and another unimportant one ($\beta_6$), including the median bias (Bias), standard deviation (SD), median standard error (SE), and coverage percentages (CP) for $(1-\alpha)100\%$ confidence intervals, where $\alpha=0.1$ and $0.05$. The coverage rates for 95\% and 90\% confidence intervals are denoted as CP95 and CP90, respectively. We observe that the debiased Lasso exhibits undercoverage, particularly when the sample size is only 200. In contrast, the confidence intervals based on the proposed DPME demonstrate coverage percentages that are reasonably close to the nominal levels, even when the sample size is small. The standard deviation and the standard error in our proposed method are in close agreement. In the last column, we list the computation time ratio of the debiased Lasso over the proposed method. Our method is computationally more efficient than debiased Lasso, especially when the dimension of covariates is larger than the sample size.

\begin{table}[!tbh]
%\centering
\caption{Simulation results based on 500 replicates for high-dimensional linear models} \label{Table_Lasso}
\begin{center}
\resizebox{\textwidth}{!}{
\begin{tabular}{cccccccccccccccc}
\toprule
    &  & & \multicolumn{5}{c}{Debiased Lasso} & &\multicolumn{5}{c}{Proposed DPME} & & Cmpt. \\ 
   \cmidrule(lr){4-8} \cmidrule(lr){10-14}
   $p$ & $n$ & & Bias & SD & SE & CP95 & CP90 & & Bias & SD & SE & CP95 & CP90 & & ratio \\
    \midrule
   100 & 200 & $\beta_1$ & 0.035 & 0.084 & 0.073 & 0.858 & 0.784 &  & -0.004 & 0.060 & 0.095 & 0.942 & 0.900 &  & 1.188 \\ 
   &  & $\beta_6$ & 0.040 & 0.065 & 0.072 & 0.936 & 0.868 &  & -0.010 & 0.084 & 0.091 & 0.966 & 0.920 &  & 1.106 \\ 
   & 500 & $\beta_1$ & 0.020 & 0.061 & 0.048 & 0.854 & 0.776 &  & 0.005 & 0.042 & 0.058 & 0.924 & 0.864 &  & 1.449 \\ 
   &  & $\beta_6$ & 0.019 & 0.051 & 0.048 & 0.918 & 0.864 &  & -0.006 & 0.060 & 0.057 & 0.954 & 0.902 &  & 1.467 \\ 
   & 1000 & $\beta_1$ & 0.002 & 0.044 & 0.035 & 0.904 & 0.834 &  & -0.005 & 0.030 & 0.041 & 0.952 & 0.888 &  & 1.769 \\ 
   &  & $\beta_6$ & 0.003 & 0.037 & 0.035 & 0.950 & 0.888 &  & -0.012 & 0.040 & 0.040 & 0.972 & 0.896 &  & 1.743 \\ 
  500 & 200 & $\beta_1$ & 0.030 & 0.098 & 0.073 & 0.854 & 0.780 &  & -0.019 & 0.070 & 0.104 & 0.960 & 0.922 &  & 18.880 \\ 
   &  & $\beta_6$ & 0.044 & 0.072 & 0.072 & 0.926 & 0.848 &  & -0.020 & 0.106 & 0.099 & 0.968 & 0.918 &  & 17.387 \\ 
   & 500 & $\beta_1$ & 0.000 & 0.057 & 0.048 & 0.890 & 0.804 &  & -0.008 & 0.039 & 0.060 & 0.948 & 0.906 &  & 10.414 \\ 
   &  & $\beta_6$ & 0.014 & 0.047 & 0.048 & 0.940 & 0.866 &  & -0.010 & 0.059 & 0.058 & 0.958 & 0.906 &  & 10.225 \\ 
   & 1000 & $\beta_1$ & -0.001 & 0.039 & 0.035 & 0.918 & 0.852 &  & -0.003 & 0.027 & 0.041 & 0.954 & 0.902 &  & 5.522 \\ 
   &  & $\beta_6$ & 0.008 & 0.033 & 0.035 & 0.944 & 0.908 &  & -0.007 & 0.036 & 0.040 & 0.976 & 0.920 &  & 6.143 \\ 
\bottomrule
\end{tabular}
} 
\end{center}
\small Note: ``SD" is standard deviation, ``SE" is standard error, ``CP95" is the coverage percentage of 95\% confidence intervals, ``CP90" is the coverage percentage of 90\% confidence intervals, and ``Cmpt. ratio" is the average ratio of computing times between the competing method and our proposed method.
\end{table}

\subsection{Simulation 2: High-Dimensional Logistic Models}

In this section, we perform a simulation study for a high-dimensional logistic model. The covariates are generated using the linear model described in Section \ref{sec_lasso}. The response variable is generated based on the logistic model,
$P(Y = 1 \mid \bX) = 1/\{1 + \exp(-\bX^\top \bbeta_0)\}$
with $\bbeta_0 = (1,1,1,1,1,0,\dots,0)^{\top}$. Our main objective is to estimate the coefficient of a significant covariate $\beta_1$ and the coefficient of an insignificant covariate $\beta_6$. To achieve this, we conduct our investigation using sample sizes of $n = 200, 500, 1000$, with each scenario replicated 500 times. The initial estimate of $\bbeta$ is calculated through logistic-Lasso using the ``glmnet'' package by selecting the tuning parameter via cross-validation. The perturbation constants $h_{1n}$ and $h_{2n}$ are the same as those in Section \ref{sec_lasso}. Furthermore, for comparison purposes, we present the results from the bias-correction inference implemented in the ``SIHR'' package \citep{cai2021optimal, guo2021inference}.

Table \ref{Table_GLM} presents the simulation results for high-dimensional logistic models after removing 2 replicates which have non-convergence issues when $n=200$. The bias-corrected method from the ``SIHR" package is unstable and shows a large bias; the standard error is severely overestimated, resulting in overcoverage of confidence intervals. %For our proposed DMPE, about 2\% of the replicates result in outliers (which deviate from the median over five standard deviations) of the estimates in the estimation of $\beta_6$, due to the singularity in calculating the inverse of the second-order numerical derivative. The table reports the results after excluding these outliers. 
For our proposed method,  the standard deviation and the standard error are in close agreement. In the last column, we list the computation time ratio of the bias-corrected method over the proposed method. Our method is generally computationally more efficient than the bias-corrected method, especially when the dimension of covariates is large compared to the sample size.

\begin{table}[!tbh]
\caption{Simulation results based on 500 replicates for high-dimensional logistic models} \label{Table_GLM}
\begin{center}
\resizebox{\textwidth}{!}{
\begin{tabular}{cccccccccccccccc}
\toprule
    &  & & \multicolumn{5}{c}{Bias-corrected} & &\multicolumn{5}{c}{Proposed DPME} & & Cmpt. \\ 
   \cmidrule(lr){4-8} \cmidrule(lr){10-14}
   $p$ & $n$ & & Bias & SD & SE & CP95 & CP90 & & Bias & SD & SE & CP95 & CP90 & & ratio \\
    \midrule
    100 & 200 & $\beta_1$ & 0.509 & 0.774 & 1.540 & 0.998 & 0.998 &  & -0.095 & 0.295 & 0.363 & 0.976 & 0.942 &  & 16.000 \\ 
    &  & $\beta_6$ & 0.036 & 0.306 & 1.206 & 1.000 & 1.000 &  & -0.038 & 0.378 & 0.383 & 0.946 & 0.900 &  & 13.591 \\ 
   & 500 & $\beta_1$ & 0.334 & 0.359 & 0.777 & 0.996 & 0.996 &  & -0.068 & 0.188 & 0.208 & 0.954 & 0.906 &  & 4.395 \\
   &  & $\beta_6$ & 0.010 & 0.172 & 0.520 & 0.994 & 0.994 &  & -0.025 & 0.198 & 0.206 & 0.948 & 0.898 &  & 3.617 \\ 
   & 1000 & $\beta_1$ & 0.337 & 0.299 & 0.677 & 0.992 & 0.990 &  & -0.048 & 0.137 & 0.144 & 0.932 & 0.890 &  & 0.973 \\ 
   &  & $\beta_6$ & -0.015 & 0.168 & 0.424 & 0.992 & 0.982 &  & -0.028 & 0.135 & 0.137 & 0.932 & 0.876 &  & 0.893 \\ 
  500 & 200 & $\beta_1$ & 0.261 & 0.480 & 0.789 & 0.994 & 0.990 &  & -0.234 & 0.266 & 0.496 & 0.978 & 0.952 &  & 8.741 \\ 
   &  & $\beta_6$ & 0.058 & 0.149 & 0.529 & 1.000 & 0.996 &  & -0.027 & 0.397 & 0.432 & 0.927 & 0.895 &  & 5.274 \\ 
   & 500 & $\beta_1$ & 1.032 & 0.879 & 1.950 & 1.000 & 0.998 &  & -0.104 & 0.198 & 0.242 & 0.956 & 0.924 &  & 5.882 \\ 
   &  & $\beta_6$ & 0.055 & 0.325 & 1.262 & 1.000 & 1.000 &  & -0.036 & 0.219 & 0.231 & 0.940 & 0.896 &  & 4.927 \\ 
   & 1000 & $\beta_1$ & 0.553 & 0.445 & 0.993 & 0.996 & 0.996 &  & -0.075 & 0.134 & 0.154 & 0.952 & 0.898 &  & 6.209 \\ 
   &  & $\beta_6$ & 0.020 & 0.247 & 0.674 & 0.998 & 0.994 &  & -0.025 & 0.139 & 0.146 & 0.924 & 0.882 &  & 5.285 \\
\bottomrule
\end{tabular}
} 
\end{center}
\small Note: See Table \ref{Table_Lasso}. 
\end{table}

\subsection{Simulation 3: Estimating Individualized Treatment Rules} \label{sec_itr}

In this section, we perform a simulation study for high-dimensional individualized treatment rules. The simulation is based on the model proposed by \citet{liang2022estimation}. Each entry of the covariates $\bX$ is generated from a standard normal distribution independently.
Let $\bbeta_\Delta = (-1,-1,1,1,0,\dots,0)^\top$, $\bbeta_S = (-1,1,-1,-1,0,\dots,0)^\top$ and $\bbeta_\pi = (0,0,1,-1,0,\dots,0)^\top$. We generate the potential outcome $Y(a) = a\Delta(\bX) + S(\bX) \varepsilon$, where $\Delta(\bX) = \bX^{\top}\bbeta_\Delta$,
$S(X) = 0.4\bX^\top\bbeta_S$, and $\pi(X;1) = 1/\{1+\exp(-0.4\bX^\top\bbeta_\pi)\}$, and $\varepsilon \sim N(0,1)$. 
We investigate sample sizes of $n = 200, 500, 1000$, with each scenario replicated 500 times. 
For identifiability, we restrict $\bbeta$ as the vector that minimizes the logistic loss with the third element $\beta_3$ standardized to 1. The true $\bbeta_0$ is approximated numerically.
We consider the decorrelated score method \citep{liang2022estimation} as a comparison. To mimic their settings, suppose that we are interested in the first eight arguments of $\bbeta$, including coefficients for four significant and four insignificant covariates. We use sample splitting and cross-fitting, as described in Section \ref{case_itr}. Following \citet{liang2022estimation}, the propensity score and outcome regression models are fitted using Gaussian kernel regression with four variables following a distance correlation-based screening, where the bandwidth is $(1.5n)^{-1/5}$. In DPME, weighted logistic-Lasso regression is performed using the ``glmnet'' package to estimate the treatment rules. The perturbation constants $h_{1n}$ and $h_{2n}$ are the same as those in Section \ref{sec_lasso}. 

\begin{table}[!tbh]
\caption{Simulation results based on 500 replicates for high-dimensional ITRs} \label{table_ITE_linear}
\begin{center}
\resizebox{\textwidth}{!}{
\begin{tabular}{cccccccccccccccc}
\toprule
$p$ & $n$ & & \multicolumn{5}{c}{Decorrelated score} & & \multicolumn{5}{c}{Proposed DPME} & & Cmpt. \\
\cmidrule(lr){4-8} \cmidrule(lr){10-14}
 & & & Bias & SD & SE & CP95 & CP90 & & Bias & SD & SE & CP95 & CP90 & & ratio \\
\midrule
  100 & 200 & $\beta_1$ & -0.003 & 0.307 & 0.156 & 0.690 & 0.601 &  & -0.007 & 0.264 & 0.376 & 0.972 & 0.956 &  & 33.482 \\ 
   &  & $\beta_6$ & -0.003 & 0.212 & 0.163 & 0.829 & 0.768 &  & -0.004 & 0.439 & 0.563 & 0.940 & 0.908 &  & 33.482 \\ 
   & 500 & $\beta_1$ & -0.022 & 0.162 & 0.083 & 0.688 & 0.620 &  & -0.014 & 0.160 & 0.178 & 0.960 & 0.928 &  & 6.427 \\ 
   &  & $\beta_6$ & -0.004 & 0.102 & 0.082 & 0.894 & 0.834 &  & -0.003 & 0.164 & 0.193 & 0.966 & 0.927 &  & 6.427 \\ 
   & 1000 & $\beta_1$ & -0.032 & 0.119 & 0.067 & 0.728 & 0.646 &  & -0.027 & 0.124 & 0.119 & 0.954 & 0.894 &  & 4.659 \\ 
   &  & $\beta_6$ & 0.006 & 0.072 & 0.064 & 0.924 & 0.866 &  & 0.004 & 0.107 & 0.110 & 0.956 & 0.912 &  & 4.659 \\ 
  500 & 200 & $\beta_1$ & -0.020 & 0.260 & 0.102 & 0.557 & 0.480 &  & -0.021 & 0.222 & 0.563 & 0.969 & 0.957 &  & 12.745 \\ 
   & & $\beta_6$ & 0.004 & 0.119 & 0.106 & 0.925 & 0.876 &  & 0.031 & 0.508 & 0.901 & 0.945 & 0.920 &  & 12.745 \\ 
   & 500 & $\beta_1$ & -0.043 & 0.164 & 0.086 & 0.685 & 0.600 &  & -0.018 & 0.150 & 0.289 & 0.970 & 0.954 &  & 18.170 \\ 
   &  & $\beta_6$ & 0.003 & 0.096 & 0.088 & 0.912 & 0.861 &  & 0.006 & 0.376 & 0.538 & 0.937 & 0.908 &  & 18.170 \\ 
   & 1000 & $\beta_1$ & -0.023 & 0.122 & 0.077 & 0.792 & 0.712 &  & -0.023 & 0.111 & 0.149 & 0.970 & 0.942 &  & 62.833 \\ 
   &  & $\beta_6$ & -0.000 & 0.093 & 0.080 & 0.910 & 0.850 &  & 0.006 & 0.157 & 0.190 & 0.940 & 0.899 &  & 62.833 \\ 
\bottomrule
\end{tabular}
} 
\end{center}
\small Note: See Table \ref{Table_Lasso}. 
\end{table}

\begin{figure}[!tbh]
\centering
\includegraphics[width=0.8\textwidth]{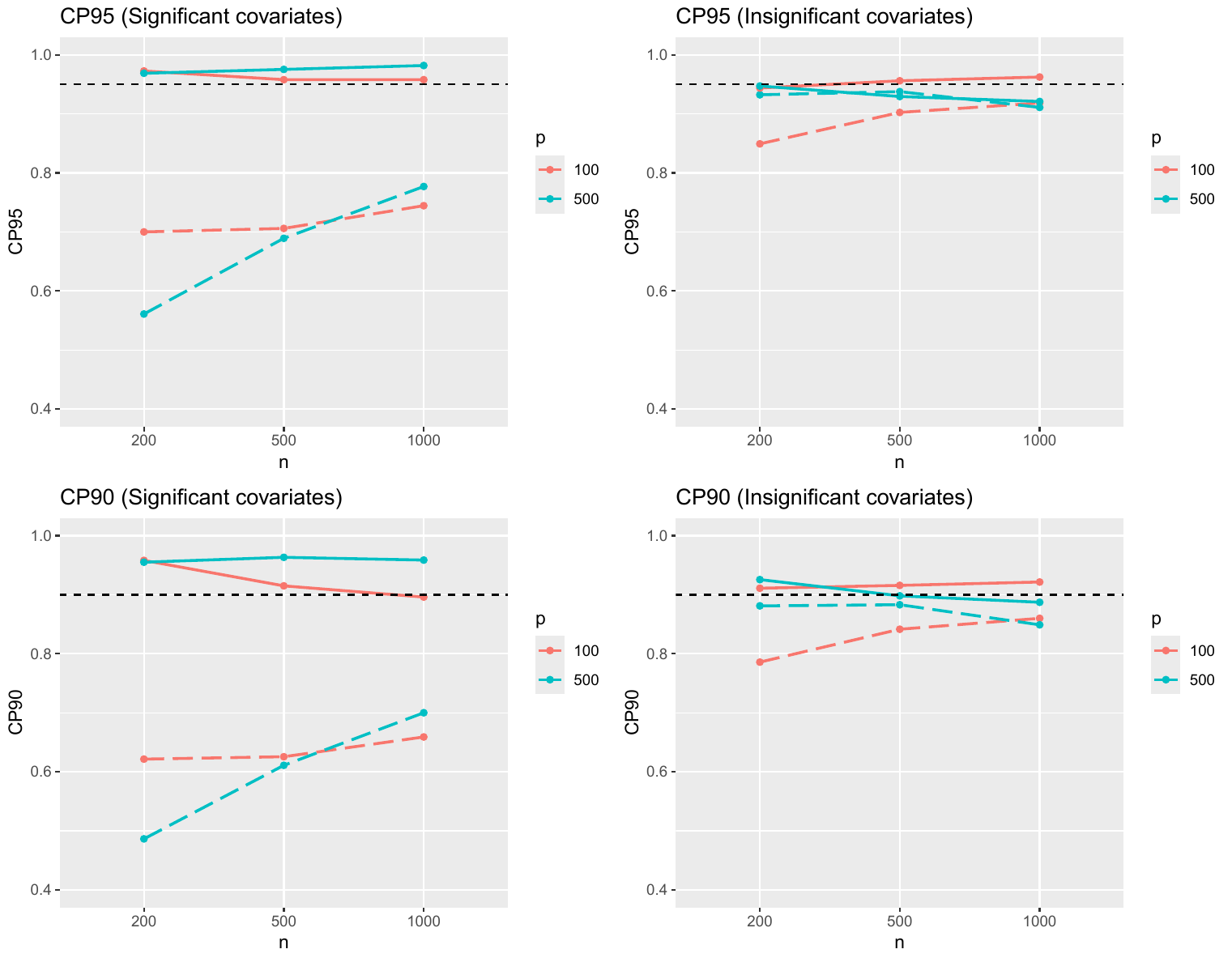}
\caption{Average coverage percentages of 95\% and 90\% confidence intervals for significant and insignificant covariates based on the decorrelated score (in dashed lines) and the proposed DPME (in solid lines) in Simulation 3.} \label{fig:CP}
\end{figure}

For the proposed method, 1.2\% of the replicates do not converge when $n=200$ and 0.2\% when $n=500$ due to the ``glmnet'' algorithm in the weighted logistic-Lasso regression for estimating the ITR. After removing the replicates with convergence issues,
we present the estimation results for the coefficient of a significant covariate $\beta_1$ and the coefficient of an insignificant covariate $\beta_6$  in Table \ref{table_ITE_linear}.
We observe that the bias is small for both methods. Figure \ref{fig:CP} presents the average coverage percentages of 95\% and 90\% confidence intervals for significant covariates ($\beta_1$--$\beta_4$) and insignificant covariates ($\beta_5$--$\beta_8$) based on the decorrelated score and the proposed DPME. The coverage rates of confidence intervals based on our proposed method are close to the nominal level for larger sample sizes, although there appears to be some overcoverage for significant coefficients, especially when $p$ is large. In contrast, the coverage rates based on the decorrelated score are much lower than the nominal level, even when $n=1000$, as this method tends to underestimate variability. In the last column, we list the computation time ratio of the decorrelated score method over our DPME in the complete inference step for all eight parameters of interest. The computation time for fitting nuisance models is not counted, as this procedure is common in both methods. The proposed method computes much faster than the decorrelated score method.

\section{Application}\label{application}

We apply our method to analyzing a clinicogenomics data set from the Gene Expression Omnibus (GEO) database (GSE9782) \citep{mulligan2007gene}. The data were collected from a study to investigate the efficacy of bortezomib in the management of multiple myeloma. Bortezomib is a targeted cancer drug that impacts gene expression, especially in pathways related to protein degradation, protein synthesis, stress, and apoptosis \citep{obeng2006proteasome} as compared to conventional chemotherapy such as dexamethasone, and has received regulatory approval for treating relapsed multiple myeloma \citep{richardson2005bortezomib, palumbo2016daratumumab}.  Our goal is to examine which specific genes influence the individual responses to bortezomib treatment.

Following the preprocessing in \citet{chen2022data}, data from two Affymetrix microarray platforms (GPL96 and GPL97) are merged to obtain a large sample size, comprising a total of 478 patients, 338 of whom received bortezomib and 140 received dexamethasone. The covariates are the standardized expression levels of 168 genes on a logarithmic scale. After removing spike-in control genes and those with a correlation larger than 0.8 with others, we finally retain 35 genes. The outcome is a response score, ranging from one (progressive disease) to five (complete response).

We use the proposed DPME to estimate a linear ITR. We randomly divide the samples into a training set and an inference set with approximately equal sizes, as described in Section \ref{sec_itr}. In the training set, we fit the propensity score and outcome regression models using kernels following variable screening. Weighted logistic-Lasso regressions are performed in the inference set to obtain the debiased estimators for all coefficients in the ITR model, where the tuning parameter is chosen through cross-validation. The final estimator is obtained by averaging over cross-fitting. To minimize the uncertainty incurred in sample splitting, we replicate the estimation procedure five times and take the median of these point estimates and standard errors. The final $P$-value for each coefficient is calculated as the harmonic mean over these repeated processes \citep{wilson2019harmonic}.
%Figure \ref{fig:qqp} presents the Q-Q plot of $P$-values for these 35 genes in the estimated ITR.

We find nine genes that are significant at 5\% level in the ITR model based on the results of DPME, as listed in Table \ref{tab:gene}. %Figure \ref{fig:qqp} shows the estimated ITR against the logarithmic expression levels of all genes for each patient. The horizontal axis is ordered based on the value of the predicted $\bX^{\top}\bbeta$ in the ITR model. The vertical axis is ordered from the most significant to the least significant gene. 
In contrast, the decorrelated score method finds 13 significant genes, possibly due to the underestimation of the standard error as seen in our simulation study. According to the estimated ITR, 301 patients are recommended for bortezomib treatment, while 177 are recommended for dexamethasone. The agreement rate between the decorrelated score and DPME is 0.79. %We also compared the estimated ITR by A-learning \citep{chen2017general}. McNamer's test indicates that there is no systemtic difference between the ITR by DPME and A-learning.

\begin{table}[!tbh]
\centering
\caption{Nine significant genes in the ITR model based on DPME} \label{tab:gene}
\begin{footnotesize}
\begin{tabular}{llcccccccc}
  \toprule
   & & & \multicolumn{3}{c}{Decorrelated score} & & \multicolumn{3}{c}{Proposed DPME} \\
  \cmidrule(lr){4-6} \cmidrule(lr){8-10} 
  Probe & Gene & & Coef & SE & $P$-value & & Coef & SE & $P$-value \\ 
  \midrule
  %200015\_s\_at & BAX & & 0.715 & 0.167 & 0.000 & & 0.801 & 0.217 & 0.000 \\ 
  %200067\_x\_at & IGKC & & -0.651 & 0.146 & 0.000 & & -0.658 & 0.183 & 0.000 \\ 
  %200039\_s\_at & PTPRC & & 0.884 & 0.181 & 0.000 & & 0.848 & 0.251 & 0.001 \\ 
  %200021\_at & EEF1B2 & & 0.430& 0.123 & 0.000 & & 0.406 & 0.161 & 0.012 \\ 
  %200023\_s\_at & RPL41 & & -0.676 & 0.180 & 0.000 & & -0.542 & 0.232 & 0.019 \\ 
  %200037\_s\_at & ANXA1 & & -0.596 & 0.207 & 0.004 & & -0.567 & 0.262 & 0.031 \\ 
  200067\_x\_at & IGKC & & -0.651 & 0.146 & 0.000 && -0.658 & 0.183 & 0.000 \\ 
  200015\_s\_at & BAX & & 0.715 & 0.167 & 0.000 && 0.801 & 0.217 & 0.000 \\ 
  200039\_s\_at & PTPRC & & 0.884 & 0.181 & 0.000 && 0.848 & 0.251 & 0.000 \\ 
  200027\_at & HSPA6 & & -0.244 & 0.192 & 0.201 && -0.513 & 0.267 & 0.000 \\ 
  200023\_s\_at & RPL41 & & -0.676 & 0.180 & 0.000 && -0.542 & 0.232 & 0.002 \\ 
  200051\_at & ATP5F1A & & -0.860 & 0.241 & 0.000 && -0.773 & 0.739 & 0.004 \\ 
  200037\_s\_at & ANXA1 & & -0.596 & 0.207 & 0.001 && -0.614 & 0.262 & 0.010 \\ 
  200021\_at & EEF1B2 & & 0.430 & 0.123 & 0.001 && 0.406 & 0.161 & 0.014 \\ 
  200085\_s\_at & CCNA2 & & -0.460 & 0.183 & 0.006 && -0.452 & 0.236 & 0.027 \\ 
   \bottomrule
\end{tabular}
\end{footnotesize}
\end{table}

%\begin{figure}[!tbh]
%\centering
%\includegraphics[width=0.4\textwidth]{qq_pvalue.pdf}
%\includegraphics[width=0.9\textwidth]{heatmap.pdf}
%\caption{The estimated ITR against 35 genes for each patient. Rows are ordered according to significance levels (most significant on the top), and columns are ordered according to the estimated $\bX^{\top}\bbeta$.}\label{fig:qqp}
%\end{figure}

Bortezomib, a proteasome inhibitor, exerts its anti-myeloma effects by disrupting protein degradation, leading to the accumulation of misfolded proteins and activation of the unfolded protein response (UPR). This stress response causes myeloma cells to curtail global protein synthesis and induce apoptosis. As a result, genes such as EEF1B2, RPL41, CCNA2, and IGKC—involved in protein synthesis, cell cycle, and plasma cell function—reflect these cellular adjustments during treatment. Meanwhile, stress- and apoptosis-related genes like BAX and HSPA6, as well as context-dependent genes such as ANXA1, PTPRC, and ATP5F1A, display expression changes consistent with the broad impact of bortezomib on cell fate and bone marrow dynamics in myeloma \citep{mitsiades2002molecular, obeng2006proteasome}.

\section{Discussion} \label{sec:conc}

Penalized estimation methods usually yield irregular estimators, making statistical inference challenging. Motivated by profile estimation and semiparametric theory, we propose a debiased profile $M$-estimator. In many scenarios, the profile $m$-function may not have an explicit form. We propose a numerical procedure that automatically approximates the debiasing direction. Through a one-step update, the resulting estimator enjoys a root-$n$ convergence rate and is asymptotically normal. DPME is an automatic profile-refitting implementation of the classical least-favorable and orthogonal correction for penalized high-dimensional $M$-estimation. In linear models, the debiasing direction of DPME coincides with the debiasing Lasso. DPME avoids solving for the explicit form of the debiasing direction, is computationally efficient, and thus applies universally to a wide range of high-dimensional settings. 
DPME is also related to the method of targeted maximum likelihood estimation in the causal inference literature \citep{van2006targeted, frangakis2015deductive, carone2019toward, ichimura2022influence,  luedtke2026simplifying}, which operationalizes profiling through a one-step update along a carefully chosen parametric submodel. However, the submodel in this method, corresponding to the least favorable submodel, must be either derived explicitly, or obtained numerically assuming that the nuisance parameters are fully nonparametric, which is not the case for the high-dimensional regression models.
%The latter will yield conservative inference and wide confidence intervals since the model space considered is much larger than the actual model space we considered in this paper.

%The proposed DPME approach has the following advantages over existing methods. First, deriving the explicit form of the submodel is avoided through numerical approximation. Here, we do not need to know the explicit form of the path-specific derivative. Second, the computation is much more efficient than the existing debiasing methods. Only a few rounds of penalized regression are required to calculate the first- and second-order numerical derivatives with respect to the target parameter. Third, the proposed DPME universally applies to high-dimensional settings, as we do not restrict the form of penalization. 

DPME offers considerable flexibility in choosing initial estimators, allowing the use of various machine learning techniques. Under mild conditions, both initial and constrained estimators can achieve an $o_p(n^{-1/4})$ convergence rate. Given a specified estimation procedure, only a few rounds of constrained optimization are needed to compute the first- and second-order numerical derivatives with respect to the target parameter.
Our framework can be used to draw inferences on low-dimensional functionals of the model parameter, such as subvectors and functions of the coefficients. In such cases, reparameterization may be necessary to facilitate constrained optimization. If the parameter of interest and nuisance parameters are not separable, constrained optimization can be achieved by involving a Lagrangian multiplier. Our approach can accommodate infinite-dimensional nuisance parameters, including those arising in partial linear models and generalized additive models.

Of note, DPME does not weaken the fundamental statistical requirements for valid high-dimensional inference. Like existing debiasing methods, it requires sufficient sparsity or regularity, local curvature, stable nuisance estimation, and a sufficiently accurate initial estimator. The advantage of DPME is computational and modular rather than information-theoretic. When an orthogonal score exists but is difficult to derive or implement analytically, DPME approximates the corresponding profile score numerically through constrained refitting. The main theorem is stated in terms of the primitive properties of the initial and constrained estimators, rather than in terms of a specific penalty. Therefore, the result can be applied to different penalties once these properties are verified. In each model, the tuning parameter must yield the required local convergence, stability of the constrained refits, and finite-difference approximation error. Modern double machine learning can bypass the ultra-sparsity condition of high-dimensional models via sample splitting. However, when the sample size is small, sample splitting may exhibit greater finite-sample variation or unstable performance.

The proposed DPME approach can also be extended to accommodate data-dependent $m$-functions. Extensions to data-dependent objectives, such as those involving first-stage nuisance estimators, require additional orthogonality and rate conditions. We illustrate one such setting in the ITR example, but a general theory for data-dependent profile objectives is beyond the scope of the present paper. Similar to the ITR estimation setting, the $m$-function may incorporate estimated nuisance parameters. For example, consider the high-dimensional Cox model, where the primary interest is in estimating the covariate coefficients $\bbeta$. If the initial estimator is obtained by maximizing the Cox partial likelihood, the $m$-function depends on the estimated baseline hazard function and is therefore data-dependent. Provided that the influence of the data-dependent component has first-order orthogonality to that of the debiasing estimator of interest, the resulting debiased estimator retains both regularity and asymptotic normality. We leave the development of a rigorous theoretical framework for data-dependent profile $m$-functions to future work.

\section*{Data Availability Statement}

The dataset analyzed in this study is publicly available from GEO under accession number [GEO: GSE9782] (https://www.ncbi.nlm.nih.gov/geo/query/acc.cgi?acc=GSE9782). The R (version 4.4.0) codes for simulation studies are part of the online supplementary material.

\section*{Competing interests}

No competing interest is declared.

%\section{Author contributions statement}

%Y.W. and Y.D. conducted the numerical studies, Y.D. analyzed the data. Y.W., Y.D., and Y.G. wrote the manuscript. Y.W. and D.Z. reviewed the manuscript.

\section*{Acknowledgments}
The authors thank the anonymous reviewers for their valuable suggestions. %This work is supported in part by funds.

\section*{Supplementary Material}

The online supplementary material includes proofs of Lemma 1 and Theorem 1.

\bibliographystyle{plainnat}
\bibliography{bibliography}



\section*{Supplementary material}

# Supplementary Material for "Debiased Inference for High-Dimensional Regression Models Based on Profile M-Estimation"

## A.  Proof of Lemma 1

By definition, $f^*(\boldsymbol{\theta})$ maximizes $Pm(\boldsymbol{Z}, f)$ under the constraint $\mathfrak{F}(f) = \boldsymbol{\theta}$. Thus, for any $h \in \mathcal{F}$ such that $\nabla\mathfrak{F}\{f^*(\boldsymbol{\theta})\}[h] = \boldsymbol{0}$, we have $P\nabla m\{\boldsymbol{Z}, f^*(\boldsymbol{\theta})\}[h] = 0$. Since Conditions 2 and 3 hold, differentiating both sides with respect to $\boldsymbol{\theta}$ yields

$$P\nabla^2 m\{\boldsymbol{Z}, f^*(\boldsymbol{\theta})\}[h, \partial_{\boldsymbol{\theta}} f^*(\boldsymbol{\theta})] = \boldsymbol{0}.$$

Plugging $\boldsymbol{\theta}_0$ into the above equations and noting that $f_0 = f^*(\boldsymbol{\theta}_0)$, we see that if $\nabla\mathfrak{F}(f_0)[h] = \langle \boldsymbol{v}^*, h \rangle = \boldsymbol{0}$, then $P\nabla^2 m(\boldsymbol{Z}, f_0)[h, \partial_{\boldsymbol{\theta}} f^*(\boldsymbol{\theta}_0)] = \boldsymbol{0}$.

For any $v \in \mathcal{F}$, let $v_h = v - \langle \boldsymbol{v}^*, v \rangle^\top \langle \boldsymbol{v}^*, \boldsymbol{v}^* \rangle^{-1} \boldsymbol{v}^*$, where $\langle \boldsymbol{v}^*, \boldsymbol{v}^* \rangle$ is a $q \times q$ matrix with the $(j, k)$th element being the inner product between the $j$th and $k$th components of $\boldsymbol{v}^*$. We can easily verify that $\langle \boldsymbol{v}^*, v_h \rangle = \boldsymbol{0}$. Thus, we have

$$\begin{aligned}
&P\nabla^2 m(\boldsymbol{Z}, f_0)[v, \partial_{\boldsymbol{\theta}} f^*(\boldsymbol{\theta}_0)] \\
&= P\nabla^2 m(\boldsymbol{Z}, f_0)[v_h, \partial_{\boldsymbol{\theta}} f^*(\boldsymbol{\theta}_0)] + P\nabla^2 m(\boldsymbol{Z}, f_0)[v - v_h, \partial_{\boldsymbol{\theta}} f^*(\boldsymbol{\theta}_0)] \\
&= \boldsymbol{0} + P\nabla^2 m(\boldsymbol{Z}, f_0)[\langle \boldsymbol{v}^*, \boldsymbol{v}^* \rangle^{-1} \boldsymbol{v}^*, \partial_{\boldsymbol{\theta}} f^*(\boldsymbol{\theta}_0)] \langle \boldsymbol{v}^*, v \rangle.
\end{aligned} \qquad (1)$$

Here, for $q$-vectors $\boldsymbol{a} = (a_1, \ldots, a_q)^\top \in \mathcal{F}^q$ and $\boldsymbol{b} = (b_1, \ldots, b_q)^\top \in \mathcal{F}^q$, the operator $P\nabla^2 m(\boldsymbol{Z}, f_0)[\boldsymbol{a}, \boldsymbol{b}]$ yields a $q \times q$ matrix whose $(j,k)$th element is $P\nabla^2 m(\boldsymbol{Z}, f_0)[a_k, b_j]$. Let $\boldsymbol{\alpha} = P\nabla^2 m(\boldsymbol{Z}, f_0)[\langle \boldsymbol{v}^*, \boldsymbol{v}^* \rangle^{-1} \boldsymbol{v}^*, \partial_{\boldsymbol{\theta}} f^*(\boldsymbol{\theta}_0)]$. Then $P\nabla^2 m(\boldsymbol{Z}, f_0)[v, \partial_{\boldsymbol{\theta}} f^*(\boldsymbol{\theta}_0)] = \boldsymbol{\alpha} \langle \boldsymbol{v}^*, v \rangle$. If $\boldsymbol{\alpha}$ is invertible, we can define

$$\boldsymbol{h}^* = \boldsymbol{\alpha}^{-1} \partial_{\boldsymbol{\theta}} f^*(\boldsymbol{\theta}_0), \tag{2}$$

so that $P\nabla^2 m(Z, f_0)[v, \boldsymbol{h}^*] = \langle \boldsymbol{v}^*, v \rangle$ for any $v \in \mathcal{F}$. In particular, we obtain

$$P\nabla^2 m(\boldsymbol{Z}, f_0)[\partial_{\boldsymbol{\theta}} f^*(\boldsymbol{\theta}_0), \boldsymbol{h}^*] = \langle \boldsymbol{v}^*, \partial_{\boldsymbol{\theta}} f^*(\boldsymbol{\theta}_0) \rangle. \tag{3}$$

According to Condition 4, it suffices to show that $\boldsymbol{\alpha} = -\boldsymbol{I}_0$. Since $\mathfrak{F}\{f^*(\boldsymbol{\theta})\} = \boldsymbol{\theta}$, differentiating both sides with respect to $\boldsymbol{\theta}$ and letting $\boldsymbol{\theta} = \boldsymbol{\theta}_0$ yield

$$\langle \boldsymbol{v}^*, \partial_{\boldsymbol{\theta}} f^*(\boldsymbol{\theta}_0) \rangle = \nabla \mathfrak{F}(f_0)[\partial_{\boldsymbol{\theta}} f^*(\boldsymbol{\theta}_0)] = \boldsymbol{1}_{q \times q}, \tag{4}$$

where $\boldsymbol{1}_{q \times q}$ denotes the $q \times q$ identity matrix. By the definition of $\boldsymbol{I}_0$, we have

$$-\boldsymbol{I}_0 = P\nabla^2 m(Z, f_0)[\partial_{\boldsymbol{\theta}} f^*(\boldsymbol{\theta}_0), \partial_{\boldsymbol{\theta}} f^*(\boldsymbol{\theta}_0)]$$

$$= \boldsymbol{\alpha} P\nabla^2 m(Z, f_0)[\partial_{\boldsymbol{\theta}} f^*(\boldsymbol{\theta}_0), \boldsymbol{h}^*]$$

$$= \boldsymbol{\alpha} \langle \boldsymbol{v}^*, \partial_{\boldsymbol{\theta}} f^*(\boldsymbol{\theta}_0) \rangle = \boldsymbol{\alpha},$$

where the second equality follows from (2), the third from (3), and the last from (4). Thus, we conclude that $\boldsymbol{h}^* = -\boldsymbol{I}_0^{-1} \partial_{\boldsymbol{\theta}} f^*(\boldsymbol{\theta}_0)$ satisfies

$$P\nabla^2 m(\boldsymbol{Z}, f_0)[\boldsymbol{h}^*, v] = \langle \boldsymbol{v}^*, v \rangle, \quad \text{for all } v \in \mathcal{F}.$$

## B. Proof of Theorem 1

First, we note that under Conditions 1 and 5,

$$\widehat{\boldsymbol{\theta}}_n - \boldsymbol{\theta}_0 = \mathfrak{F}(\widehat{f}_n) - \mathfrak{F}(f_0) = \langle \boldsymbol{v}^*, \widehat{f}_n - f_0 \rangle + o_p\{d(\widehat{f}_n, f_0)\} = o_p(n^{-1/4}).$$

By the construction of $\widetilde{\boldsymbol{\theta}}_n$, we have

$$\sqrt{n}(\widetilde{\boldsymbol{\theta}}_n - \boldsymbol{\theta}_0) = \sqrt{n}(\widehat{\boldsymbol{\theta}}_n - \boldsymbol{\theta}_0) - \sqrt{n}\left\{D^2_{h_{2n}} A_n(\widehat{\boldsymbol{\theta}}_n)\right\}^{-1} D_{h_{1n}} A_n(\widehat{\boldsymbol{\theta}}_n).$$

By Condition 1 and Lemma 1, $\widehat{\boldsymbol{\theta}}_n - \boldsymbol{\theta}_0$ can also be written as

$$\begin{aligned}
&\mathfrak{F}\{f^*(\widehat{\boldsymbol{\theta}}_n)\} - \mathfrak{F}\{f^*(\boldsymbol{\theta}_0)\} \\
&= \left\langle \boldsymbol{v}^*, f^*(\widehat{\boldsymbol{\theta}}_n) - f^*(\boldsymbol{\theta}_0)\right\rangle + O_p\left[d^2\left\{f^*(\widehat{\boldsymbol{\theta}}_n), f^*(\boldsymbol{\theta}_0)\right\}\right] \\
&= P\nabla^2 m(\boldsymbol{Z}, f_0)[\boldsymbol{h}^*, f^*(\widehat{\boldsymbol{\theta}}_n) - f^*(\boldsymbol{\theta}_0)] + o_p(n^{-1/2}) \\
&= \left[P\partial^2_{\boldsymbol{\theta}\boldsymbol{\theta}} m\{\boldsymbol{Z}, f^*(\boldsymbol{\theta}_0)\}\right]^{-1} P\nabla^2 m\{\boldsymbol{Z}, f^*(\boldsymbol{\theta}_0)\}[\partial_{\boldsymbol{\theta}} f^*(\boldsymbol{\theta}_0), f^*(\widehat{\boldsymbol{\theta}}_n) - f^*(\boldsymbol{\theta}_0)] + o_p(n^{-1/2}).
\end{aligned}$$

Next, we will show that using numerical derivatives to approximate true derivatives only brings error up to $o_p(n^{-1/2})$. Note that

$$\begin{aligned}
0 &= \left[\mathfrak{F}\{\widehat{f}_n(\widehat{\boldsymbol{\theta}}_n)\} - \mathfrak{F}(f_0)\right] - \left[\mathfrak{F}\{f^*(\widehat{\boldsymbol{\theta}}_n)\} - \mathfrak{F}(f_0)\right] \\
&= \nabla\mathfrak{F}(f_0)[\widehat{f}_n(\widehat{\boldsymbol{\theta}}_n) - f_0] + O_p\left[d^2\left\{\widehat{f}_n(\widehat{\boldsymbol{\theta}}_n), f_0\right\}\right] \\
&\quad - \nabla\mathfrak{F}(f_0)[f^*(\widehat{\boldsymbol{\theta}}_n) - f_0] + O_p\left[d^2\left\{f^*(\widehat{\boldsymbol{\theta}}_n), f_0\right\}\right] \\
&= \nabla\mathfrak{F}(f_0)[\widehat{f}_n(\widehat{\boldsymbol{\theta}}_n) - f^*(\widehat{\boldsymbol{\theta}}_n)] + o_p(n^{-1/2}),
\end{aligned}$$

so

$$\begin{aligned}
&P\nabla^2 m\{\boldsymbol{Z}, f^*(\boldsymbol{\theta}_0)\}[\partial_{\boldsymbol{\theta}} f^*(\boldsymbol{\theta}_0), \widehat{f}_n(\widehat{\boldsymbol{\theta}}_n) - f^*(\widehat{\boldsymbol{\theta}}_n)] \\
&= -\boldsymbol{I}_0 P\nabla^2 m\{\boldsymbol{Z}, f^*(\boldsymbol{\theta}_0)\}[\boldsymbol{h}^*, \widehat{f}_n(\widehat{\boldsymbol{\theta}}_n) - f^*(\widehat{\boldsymbol{\theta}}_n)] \\
&= -\boldsymbol{I}_0 \nabla\mathfrak{F}(f_0)[\widehat{f}_n(\widehat{\boldsymbol{\theta}}_n) - f^*(\widehat{\boldsymbol{\theta}}_n)] \\
&= o_p(n^{-1/2}).
\end{aligned}$$

Now we establish the error of approximating the first-order derivative. Under Conditions

7 and 8,

$$D_{h_{1n}}A(\widehat{\boldsymbol{\theta}}_n) = \mathbb{P}_n D_{h_{1n}} m\{\boldsymbol{Z}, \widehat{f}_n(\widehat{\boldsymbol{\theta}}_n)\}$$

$$= (\mathbb{P}_n - P)D_{h_{1n}} m\{\boldsymbol{Z}, \widehat{f}_n(\widehat{\boldsymbol{\theta}}_n)\} + PD_{h_{1n}} m\{\boldsymbol{Z}, \widehat{f}_n(\widehat{\boldsymbol{\theta}}_n)\}$$

$$= (\mathbb{P}_n - P)\partial_{\boldsymbol{\theta}} m\{\boldsymbol{Z}, f^*(\widehat{\boldsymbol{\theta}}_n)\} + PD_{h_{1n}} m\{\boldsymbol{Z}, \widehat{f}_n(\widehat{\boldsymbol{\theta}}_n)\} + o_p(n^{-1/2})$$

$$= \mathbb{P}_n \partial_{\boldsymbol{\theta}} m\{\boldsymbol{Z}, f^*(\widehat{\boldsymbol{\theta}}_n)\} + PD_{h_{1n}} m\{\boldsymbol{Z}, \widehat{f}_n(\widehat{\boldsymbol{\theta}}_n)\} + o_p(n^{-1/2}).$$

For the second term in the last equation,

$$PD_{h_{1n}} m\{\boldsymbol{Z}, \widehat{f}_n(\widehat{\boldsymbol{\theta}}_n)\} = \frac{1}{h_{1n}}\left[Pm\{\boldsymbol{Z}, \widehat{f}_n(\widehat{\boldsymbol{\theta}}_n + h_{1n})\} - Pm\{\boldsymbol{Z}, \widehat{f}_n(\widehat{\boldsymbol{\theta}}_n)\}\right]$$

$$= \frac{1}{h_{1n}}\Big(Pm\{\boldsymbol{Z}, f^*(\boldsymbol{\theta}_0 + h_{1n})\} - Pm\{\boldsymbol{Z}, f^*(\boldsymbol{\theta}_0)\}$$

$$+ P\nabla m\{\boldsymbol{Z}, f^*(\boldsymbol{\theta}_0 + h_{1n})\}[\widehat{f}_n(\widehat{\boldsymbol{\theta}}_n + h_{1n}) - f^*(\boldsymbol{\theta}_0 + h_{1n})]$$

$$- P\nabla m\{\boldsymbol{Z}, f^*(\boldsymbol{\theta}_0)\}[\widehat{f}_n(\widehat{\boldsymbol{\theta}}_n) - f^*(\boldsymbol{\theta}_0)]\Big)$$

$$= P\partial_{\boldsymbol{\theta}} m\{\boldsymbol{Z}, f^*(\boldsymbol{\theta}_0)\} + O(h_{1n})$$

$$+ \frac{1}{h_{1n}} P\nabla m\{\boldsymbol{Z}, f^*(\boldsymbol{\theta}_0)\}[f^*(\boldsymbol{\theta}_0 + h_{1n}) - f^*(\boldsymbol{\theta}_0), \widehat{f}_n(\widehat{\boldsymbol{\theta}}_n) - f^*(\widehat{\boldsymbol{\theta}}_n)]$$

$$+ \frac{1}{h_{1n}} O_p\left(d^2\{f^*(\widehat{\boldsymbol{\theta}}_n + h_{1n}), f^*(\widehat{\boldsymbol{\theta}}_n)\}\right)$$

$$= P\nabla m\{\boldsymbol{Z}, f^*(\boldsymbol{\theta}_0)\}[\partial_{\boldsymbol{\theta}} f^*(\boldsymbol{\theta}_0) + O(h_{1n}), \widehat{f}_n(\widehat{\boldsymbol{\theta}}_n) - f^*(\widehat{\boldsymbol{\theta}}_n)] + O_p(h_{1n})$$

$$= P\nabla m\{\boldsymbol{Z}, f^*(\boldsymbol{\theta}_0)\}[\partial_{\boldsymbol{\theta}} f^*(\boldsymbol{\theta}_0), \widehat{f}_n(\widehat{\boldsymbol{\theta}}_n) - f^*(\widehat{\boldsymbol{\theta}}_n)]$$

$$+ O_p\left(h_{1n} d\{\widehat{f}_n(\widehat{\boldsymbol{\theta}}_n), f^*(\widehat{\boldsymbol{\theta}}_n)\}\right) + O_p(h_{1n})$$

$$= o_p(n^{-1/2}) + O_p(h_{1n}).$$

Therefore,

$$D_{h_{1n}}A(\widehat{\boldsymbol{\theta}}_n) = \mathbb{P}_n \partial_{\boldsymbol{\theta}} m\{\boldsymbol{Z}, f^*(\widehat{\boldsymbol{\theta}}_n)\} + o_p(n^{-1/2}) + O_p(h_{1n}).$$

Given $h_{1n} = o(n^{-1/2})$, we have

$$D_{h_{1n}}A(\widehat{\boldsymbol{\theta}}_n) = \mathbb{P}_n \partial_{\boldsymbol{\theta}} m\{\boldsymbol{Z}, f^*(\widehat{\boldsymbol{\theta}}_n)\} + o_p(n^{-1/2}).$$

Now we establish the error of approximating the second-order derivative. Under Conditions 7 and 8,

$$D^2_{h_{2n}} A(\widehat{\boldsymbol{\theta}}_n) = \mathbb{P}_n D^2_{h_{2n}} m\{\boldsymbol{Z}, \widehat{f}_n(\widehat{\boldsymbol{\theta}}_n)\}$$

$$= (\mathbb{P}_n - P) D^2_{h_{2n}} m\{\boldsymbol{Z}, \widehat{f}_n(\widehat{\boldsymbol{\theta}}_n)\} + P D^2_{h_{2n}} m\{\boldsymbol{Z}, \widehat{f}_n(\widehat{\boldsymbol{\theta}}_n)\}$$

$$= (\mathbb{P}_n - P) \partial_{\boldsymbol{\theta}\boldsymbol{\theta}} m\{\boldsymbol{Z}, f^*(\widehat{\boldsymbol{\theta}}_n)\} + P D^2_{h_{2n}} m\{\boldsymbol{Z}, \widehat{f}_n(\widehat{\boldsymbol{\theta}}_n)\} + o_p(n^{-1/2})$$

$$= O_p(n^{-1/2}) + P D^2_{h_{2n}} m\{\boldsymbol{Z}, \widehat{f}_n(\widehat{\boldsymbol{\theta}}_n)\}.$$

For the second term in the last equation,

$$P D^2_{h_{2n}} m\{\boldsymbol{Z}, \widehat{f}_n(\widehat{\boldsymbol{\theta}}_n)\} = P D^2_{h_{2n}} m\{\boldsymbol{Z}, f^*(\widehat{\boldsymbol{\theta}}_n)\} + O_p\left(d\{\widehat{f}_n(\widehat{\boldsymbol{\theta}}_n), f^*(\widehat{\boldsymbol{\theta}}_n)\}\right)$$

$$= P \partial_{\boldsymbol{\theta}\boldsymbol{\theta}} m\{\boldsymbol{Z}, f^*(\widehat{\boldsymbol{\theta}}_n)\} + O_p(h_{2n}) + o_p(n^{-1/4}).$$

Therefore,

$$D^2_{h_{2n}} A(\widehat{\boldsymbol{\theta}}_n) = P \partial_{\boldsymbol{\theta}\boldsymbol{\theta}} m\{\boldsymbol{Z}, f^*(\widehat{\boldsymbol{\theta}}_n)\} + o_p(n^{-1/4}) + O_p(h_{2n}).$$

Given $h_{2n} = O(n^{-1/4})$, we have

$$D^2_{h_{2n}} A(\widehat{\boldsymbol{\theta}}_n) = P \partial_{\boldsymbol{\theta}\boldsymbol{\theta}} m\{\boldsymbol{Z}, f^*(\widehat{\boldsymbol{\theta}}_n)\} + O_p(n^{-1/4}).$$

Moreover, by recalling that $\mathbb{P}_n \partial_{\boldsymbol{\theta}} m\{\boldsymbol{Z}, f^*(\widehat{\boldsymbol{\theta}}_n)\} = o_p(n^{-1/4})$,

$$\left\{D^2_{h_{2n}} A_n(\widehat{\boldsymbol{\theta}}_n)\right\}^{-1} D_{h_{1n}} A_n(\widehat{\boldsymbol{\theta}}_n)$$
$$= \left[P\partial^2_{\boldsymbol{\theta\theta}} m\{\boldsymbol{Z}, f^*(\widehat{\boldsymbol{\theta}}_n)\} + O_p(n^{-1/4})\right]^{-1} \left[\mathbb{P}_n \partial_{\boldsymbol{\theta}} m\{\boldsymbol{Z}, f^*(\widehat{\boldsymbol{\theta}}_n)\} + o_p(n^{-1/2})\right]$$
$$= \left[P\partial^2_{\boldsymbol{\theta\theta}} m\{\boldsymbol{Z}, f^*(\widehat{\boldsymbol{\theta}}_n)\}\right]^{-1} \mathbb{P}_n \partial_{\boldsymbol{\theta}} m\{\boldsymbol{Z}, f^*(\widehat{\boldsymbol{\theta}}_n)\} + o_p(n^{-1/2})$$
$$= \left[P\partial^2_{\boldsymbol{\theta\theta}} m\{\boldsymbol{Z}, f^*(\widehat{\boldsymbol{\theta}}_n)\}\right]^{-1} (\mathbb{P}_n - P)\partial_{\boldsymbol{\theta}} m\{\boldsymbol{Z}, f^*(\widehat{\boldsymbol{\theta}}_n)\}$$
$$+ \left[P\partial^2_{\boldsymbol{\theta\theta}} m\{\boldsymbol{Z}, f^*(\widehat{\boldsymbol{\theta}}_n)\}\right]^{-1} P\left[\partial_{\boldsymbol{\theta}} m\{\boldsymbol{Z}, f^*(\widehat{\boldsymbol{\theta}}_n)\} - \partial_{\boldsymbol{\theta}} m\{\boldsymbol{Z}, f^*(\boldsymbol{\theta}_0)\}\right] + o_p(n^{-1/2})$$
$$= n^{-1/2} \left[P\partial^2_{\boldsymbol{\theta\theta}} m\{\boldsymbol{Z}, f^*(\widehat{\boldsymbol{\theta}}_n)\}\right]^{-1} \mathbb{G}_n \partial_{\boldsymbol{\theta}} m\{\boldsymbol{Z}, f^*(\widehat{\boldsymbol{\theta}}_n)\}$$
$$+ \left[P\partial^2_{\boldsymbol{\theta\theta}} m\{\boldsymbol{Z}, f^*(\widehat{\boldsymbol{\theta}}_n)\}\right]^{-1} P\nabla^2 m\{\boldsymbol{Z}, f^*(\boldsymbol{\theta}_0)\}[\partial_{\boldsymbol{\theta}} f^*(\boldsymbol{\theta}_0), f^*(\widehat{\boldsymbol{\theta}}_n) - f^*(\boldsymbol{\theta}_0)]$$
$$+ O_p(\|\widehat{\boldsymbol{\theta}}_n - \boldsymbol{\theta}_0\|^2) + o_p(n^{-1/2})$$
$$= n^{-1/2} \left[P\partial^2_{\boldsymbol{\theta\theta}} m\{\boldsymbol{Z}, f^*(\widehat{\boldsymbol{\theta}}_n)\}\right]^{-1} \mathbb{G}_n \partial_{\boldsymbol{\theta}} m\{\boldsymbol{Z}, f^*(\widehat{\boldsymbol{\theta}}_n)\}$$
$$+ \left[P\partial^2_{\boldsymbol{\theta\theta}} m\{\boldsymbol{Z}, f^*(\widehat{\boldsymbol{\theta}}_n)\}\right]^{-1} P\nabla^2 m\{\boldsymbol{Z}, f^*(\boldsymbol{\theta}_0)\}[\partial_{\boldsymbol{\theta}} f^*(\boldsymbol{\theta}_0), f^*(\widehat{\boldsymbol{\theta}}_n) - f^*(\boldsymbol{\theta}_0)] + o_p(n^{-1/2}).$$

Combining the above results, we obtain that

$$\sqrt{n}(\widetilde{\boldsymbol{\theta}}_n - \boldsymbol{\theta}_0)$$
$$= \sqrt{n}\left[P\partial^2_{\boldsymbol{\theta\theta}} m\{\boldsymbol{Z}, f^*(\boldsymbol{\theta}_0)\}\right]^{-1} P\nabla^2 m\{\boldsymbol{Z}, f^*(\boldsymbol{\theta}_0)\}[\partial_{\boldsymbol{\theta}} f^*(\boldsymbol{\theta}_0), f^*(\widehat{\boldsymbol{\theta}}_n) - f^*(\boldsymbol{\theta}_0)] + o_p(1)$$
$$- \left[P\partial^2_{\boldsymbol{\theta\theta}} m\{\boldsymbol{Z}, f^*(\widehat{\boldsymbol{\theta}}_n)\}\right]^{-1} \mathbb{G}_n \partial_{\boldsymbol{\theta}} m\{\boldsymbol{Z}, f^*(\widehat{\boldsymbol{\theta}}_n)\}$$
$$- \sqrt{n}\left[P\partial^2_{\boldsymbol{\theta\theta}} m\{\boldsymbol{Z}, f^*(\widehat{\boldsymbol{\theta}}_n)\}\right]^{-1} P\nabla^2 m\{\boldsymbol{Z}, f^*(\boldsymbol{\theta}_0)\}[\partial_{\boldsymbol{\theta}} f^*(\boldsymbol{\theta}_0), f^*(\widehat{\boldsymbol{\theta}}_n) - f^*(\boldsymbol{\theta}_0)] + o_p(1)$$
$$= -\left[P\partial^2_{\boldsymbol{\theta\theta}} m\{\boldsymbol{Z}, f^*(\widehat{\boldsymbol{\theta}}_n)\}\right]^{-1} \mathbb{G}_n \partial_{\boldsymbol{\theta}} m\{\boldsymbol{Z}, f^*(\widehat{\boldsymbol{\theta}}_n)\}$$
$$- \sqrt{n}\left(\left[P\partial^2_{\boldsymbol{\theta\theta}} m\{\boldsymbol{Z}, f^*(\widehat{\boldsymbol{\theta}}_n)\}\right]^{-1} - \left[P\partial^2_{\boldsymbol{\theta\theta}} m\{\boldsymbol{Z}, f^*(\boldsymbol{\theta}_0)\}\right]^{-1}\right)$$
$$\times P\nabla^2 m\{\boldsymbol{Z}, f^*(\boldsymbol{\theta}_0)\}[\partial_{\boldsymbol{\theta}} f^*(\boldsymbol{\theta}_0), f^*(\widehat{\boldsymbol{\theta}}_n) - f^*(\boldsymbol{\theta}_0)] + o_p(1).$$

(5)

By Condition 6 and Lemma 1, the first term on the right-hand side satisfies

$$-\left[P\partial^2_{\boldsymbol{\theta\theta}}m\{\boldsymbol{Z},f^*(\widehat{\boldsymbol{\theta}}_n)\}\right]^{-1}\mathbb{G}_n\partial_{\boldsymbol{\theta}}m\{\boldsymbol{Z},f^*(\widehat{\boldsymbol{\theta}}_n)\}$$

$$=-\left[P\partial^2_{\boldsymbol{\theta\theta}}m\{\boldsymbol{Z},f^*(\widehat{\boldsymbol{\theta}}_n)\}\right]^{-1}\mathbb{G}_n\nabla m\{\boldsymbol{Z},f^*(\widehat{\boldsymbol{\theta}}_n)\}[\partial_{\boldsymbol{\theta}}f^*(\widehat{\boldsymbol{\theta}}_n)]$$

$$=-\left[P\partial^2_{\boldsymbol{\theta\theta}}m\{\boldsymbol{Z},f^*(\boldsymbol{\theta}_0)\}\right]^{-1}\mathbb{G}_n\nabla m\{\boldsymbol{Z},f^*(\boldsymbol{\theta}_0)\}[\partial_{\boldsymbol{\theta}}f^*(\boldsymbol{\theta}_0)] + o_p(1)$$

$$=-\mathbb{G}_n\nabla m(\boldsymbol{Z},f_0)[\boldsymbol{h}^*] + o_p(1).$$

In addition, since the matrix $P\partial^2_{\boldsymbol{\theta\theta}}m\{\boldsymbol{Z},f^*(\widehat{\boldsymbol{\theta}}_n)\}$ converges to $-\boldsymbol{I}_0$, it is invertible. Thus, the second term on the right-hand side of (5) satisfies

$$-\sqrt{n}\left(\left[P\partial^2_{\boldsymbol{\theta\theta}}m\{\boldsymbol{Z},f^*(\widehat{\boldsymbol{\theta}}_n)\}\right]^{-1} - \left[P\partial^2_{\boldsymbol{\theta\theta}}m\{\boldsymbol{Z},f^*(\boldsymbol{\theta}_0)\}\right]^{-1}\right)$$

$$\times P\nabla^2 m\{\boldsymbol{Z},f^*(\boldsymbol{\theta}_0)\}[\partial_{\boldsymbol{\theta}}f^*(\boldsymbol{\theta}_0),f^*(\widehat{\boldsymbol{\theta}}_n)-f^*(\boldsymbol{\theta}_0)]$$

$$=-\sqrt{n}\times O_p(\|\widehat{\boldsymbol{\theta}}_n-\boldsymbol{\theta}_0\|)\times O_p(\|\widehat{\boldsymbol{\theta}}_n-\boldsymbol{\theta}_0\|)$$

$$=o_p(1).$$

Finally, we conclude that

$$\sqrt{n}(\widetilde{\boldsymbol{\theta}}_n-\boldsymbol{\theta}_0) = -\mathbb{G}_n\nabla m(\boldsymbol{Z},f_0)[\boldsymbol{h}^*] + o_p(1).$$

## C. Proof of Sufficiency of Condition 8'

Suppose that $\widehat{f}_n(\boldsymbol{\theta})$ is differentiable with respect to $\boldsymbol{\theta}$, and

$$(\mathbb{P}_n - P)\nabla^2 m(\boldsymbol{Z},f)[v_1,v_2] = O_p(n^{-1/4}) \tag{6}$$

in the neighborhood of $f_0$, $\partial f^*(\boldsymbol{\theta}_0)$, and $\partial f^*(\boldsymbol{\theta}_0)$ for $f$, $v_1$, and $v_2$, respectively. Given $h_{1n} = o_p(n^{-1/2})$ and $h_{2n} = O_p(n^{-1/4})$, we will show that Condition 8' ensures

$$D_{h_{1n}}A(\widehat{\boldsymbol{\theta}}_n) = \mathbb{P}_n\partial_{\boldsymbol{\theta}}m\{\boldsymbol{Z},f^*(\widehat{\boldsymbol{\theta}}_n)\} + o_p(n^{-1/2}), \tag{7}$$

$$D^2_{h_{2n}}A(\widehat{\boldsymbol{\theta}}_n) = P\partial^2_{\boldsymbol{\theta\theta}}m\{\boldsymbol{Z},f^*(\widehat{\boldsymbol{\theta}}_n)\} + O_p(n^{-1/4}). \tag{8}$$

By the Taylor expansion, we have

$$D_{h_{1n}} A_n(\widehat{\boldsymbol{\theta}}_n) = \mathbb{P}_n \partial_{\boldsymbol{\theta}} m\{\boldsymbol{Z}, \widehat{f}_n(\widehat{\boldsymbol{\theta}}_n)\} + O_p(h_{1n}),$$

$$D^2_{h_{2n}} A_n(\widehat{\boldsymbol{\theta}}_n) = \mathbb{P}_n \partial^2_{\boldsymbol{\theta}\boldsymbol{\theta}} m\{\boldsymbol{Z}, \widehat{f}_n(\widehat{\boldsymbol{\theta}}_n)\} + O_p(h_{2n}).$$

To show (7), note that under Condition 5,

$$D_{h_{1n}} A_n(\widehat{\boldsymbol{\theta}}_n) - \mathbb{P}_n \partial_{\boldsymbol{\theta}} m\{\boldsymbol{Z}, f^*(\widehat{\boldsymbol{\theta}}_n)\}$$

$$= \mathbb{P}_n \partial_{\boldsymbol{\theta}} m\{\boldsymbol{Z}, \widehat{f}_n(\widehat{\boldsymbol{\theta}}_n)\} - \mathbb{P}_n \partial_{\boldsymbol{\theta}} m\{\boldsymbol{Z}, f^*(\widehat{\boldsymbol{\theta}}_n)\} + O_p(h_{1n})$$

$$= (\mathbb{P}_n - P)\left(\nabla m\{\boldsymbol{Z}, \widehat{f}_n(\widehat{\boldsymbol{\theta}}_n)\}[\partial_{\boldsymbol{\theta}} \widehat{f}_n(\widehat{\boldsymbol{\theta}}_n)] - \nabla m\{\boldsymbol{Z}, f^*(\widehat{\boldsymbol{\theta}}_n)\}[\partial_{\boldsymbol{\theta}} f^*(\widehat{\boldsymbol{\theta}}_n)]\right)$$

$$+ P\left(\nabla m\{\boldsymbol{Z}, \widehat{f}_n(\widehat{\boldsymbol{\theta}}_n)\}[\partial_{\boldsymbol{\theta}} \widehat{f}_n(\widehat{\boldsymbol{\theta}}_n)] - \nabla m\{\boldsymbol{Z}, f^*(\widehat{\boldsymbol{\theta}}_n)\}[\partial_{\boldsymbol{\theta}} f^*(\widehat{\boldsymbol{\theta}}_n)]\right) + O_p(h_{1n})$$

$$= n^{-1/2} \mathbb{G}_n \left(\nabla m\{\boldsymbol{Z}, \widehat{f}_n(\widehat{\boldsymbol{\theta}}_n)\}[\partial_{\boldsymbol{\theta}} \widehat{f}_n(\widehat{\boldsymbol{\theta}}_n)] - \nabla m\{\boldsymbol{Z}, f^*(\widehat{\boldsymbol{\theta}}_n)\}[\partial_{\boldsymbol{\theta}} f^*(\widehat{\boldsymbol{\theta}}_n)]\right)$$

$$+ P\left(\nabla m\{\boldsymbol{Z}, \widehat{f}_n(\widehat{\boldsymbol{\theta}}_n)\}[\partial_{\boldsymbol{\theta}} \widehat{f}_n(\widehat{\boldsymbol{\theta}}_n) - \partial_{\boldsymbol{\theta}} f^*(\widehat{\boldsymbol{\theta}}_n)] - \nabla m\{\boldsymbol{Z}, f^*(\boldsymbol{\theta}_0)\}[\partial_{\boldsymbol{\theta}} \widehat{f}_n(\widehat{\boldsymbol{\theta}}_n) - \partial_{\boldsymbol{\theta}} f^*(\widehat{\boldsymbol{\theta}}_n)]\right)$$

$$+ P\nabla^2 m\{\boldsymbol{Z}, f^*(\widehat{\boldsymbol{\theta}}_n)\}[\partial_{\boldsymbol{\theta}} f^*(\widehat{\boldsymbol{\theta}}_n), \widehat{f}_n(\widehat{\boldsymbol{\theta}}_n) - f^*(\widehat{\boldsymbol{\theta}}_n)] + O_p\left[d^2\left\{\widehat{f}_n(\widehat{\boldsymbol{\theta}}_n), f^*(\widehat{\boldsymbol{\theta}}_n)\right\}\right] + O_p(h_{1n})$$

$$= n^{-1/2} \mathbb{G}_n \left(\nabla m\{\boldsymbol{Z}, \widehat{f}_n(\widehat{\boldsymbol{\theta}}_n)\}[\partial_{\boldsymbol{\theta}} \widehat{f}_n(\widehat{\boldsymbol{\theta}}_n)] - \nabla m\{\boldsymbol{Z}, f^*(\widehat{\boldsymbol{\theta}}_n)\}[\partial_{\boldsymbol{\theta}} f^*(\widehat{\boldsymbol{\theta}}_n)]\right)$$

$$+ P\nabla^2 m\{\boldsymbol{Z}, f^*(\boldsymbol{\theta}_0)\}[\partial_{\boldsymbol{\theta}} \widehat{f}_n(\widehat{\boldsymbol{\theta}}_n) - \partial_{\boldsymbol{\theta}} f^*(\widehat{\boldsymbol{\theta}}_n), \widehat{f}_n(\widehat{\boldsymbol{\theta}}_n) - f^*(\boldsymbol{\theta}_0)]$$

$$+ P\nabla^2 m\{\boldsymbol{Z}, f^*(\widehat{\boldsymbol{\theta}}_n)\}[\partial_{\boldsymbol{\theta}} f^*(\widehat{\boldsymbol{\theta}}_n), \widehat{f}_n(\widehat{\boldsymbol{\theta}}_n) - f^*(\widehat{\boldsymbol{\theta}}_n)] + o_p(n^{-1/2}) + O_p(h_{1n}), \tag{9}$$

where the last inequality follows from Condition 5 and the fact that

$$d\{\widehat{f}_n(\widehat{\boldsymbol{\theta}}_n), f^*(\widehat{\boldsymbol{\theta}}_n)\} = d\{\widehat{f}_n, f^*(\widehat{\boldsymbol{\theta}}_n)\} \leq d(\widehat{f}_n, f_0) + d\{f^*(\boldsymbol{\theta}_0), f^*(\widehat{\boldsymbol{\theta}}_n)\} = o_p(n^{-1/4}).$$

By Conditions 6 and 8', the first and second terms on the right-hand side of (9) are both

$o_p(n^{-1/2})$. The third term on the right-hand side of (9) can be further written as

$$P\left(\nabla^2 m\{\boldsymbol{Z}, f^*(\widehat{\boldsymbol{\theta}}_n)\}[\partial_{\boldsymbol{\theta}} f^*(\widehat{\boldsymbol{\theta}}_n), \widehat{f}_n(\widehat{\boldsymbol{\theta}}_n) - f^*(\widehat{\boldsymbol{\theta}}_n)]\right.$$
$$\left.-\nabla^2 m\{\boldsymbol{Z}, f^*(\boldsymbol{\theta}_0)\}[\partial_{\boldsymbol{\theta}} f^*(\boldsymbol{\theta}_0), \widehat{f}_n(\widehat{\boldsymbol{\theta}}_n) - f^*(\widehat{\boldsymbol{\theta}}_n)]\right)$$
$$+ P\nabla^2 m\{\boldsymbol{Z}, f^*(\boldsymbol{\theta}_0)\}[\partial_{\boldsymbol{\theta}} f^*(\boldsymbol{\theta}_0), \widehat{f}_n(\widehat{\boldsymbol{\theta}}_n) - f^*(\widehat{\boldsymbol{\theta}}_n)]$$
$$= O_p(\|\widehat{\boldsymbol{\theta}}_n - \boldsymbol{\theta}_0\|) \times O_p\left[d\left\{\widehat{f}_n(\widehat{\boldsymbol{\theta}}_n), f^*(\widehat{\boldsymbol{\theta}}_n)\right\}\right] + o_p(n^{-1/2})$$
$$= o_p(n^{-1/2}).$$

Applying the above results to the right-hand side of (9), we obtain

$$D_{h_{1n}} A_n(\widehat{\boldsymbol{\theta}}_n) - \mathbb{P}_n \partial_{\boldsymbol{\theta}} m\{\boldsymbol{Z}, f^*(\widehat{\boldsymbol{\theta}}_n)\} = o_p(n^{-1/2}) + O_p(h_{1n}),$$

which entails (7) when $h_{1n} = o_p(n^{-1/2})$.

To show (8), note that under Condition 8',

$$D_{h_{2n}}^2 A_n(\widehat{\boldsymbol{\theta}}_n) - P\partial_{\boldsymbol{\theta}\boldsymbol{\theta}}^2 m\{\boldsymbol{Z}, f^*(\widehat{\boldsymbol{\theta}}_n)\}$$
$$= \mathbb{P}_n \partial_{\boldsymbol{\theta}\boldsymbol{\theta}}^2 m\{\boldsymbol{Z}, \widehat{f}_n(\widehat{\boldsymbol{\theta}}_n)\} - P\partial_{\boldsymbol{\theta}\boldsymbol{\theta}}^2 m\{\boldsymbol{Z}, f^*(\widehat{\boldsymbol{\theta}}_n)\} + O_p(h_{2n})$$
$$= (\mathbb{P}_n - P)\nabla^2 m\{\boldsymbol{Z}, \widehat{f}_n(\widehat{\boldsymbol{\theta}}_n)\}[\partial_{\boldsymbol{\theta}}\widehat{f}_n(\widehat{\boldsymbol{\theta}}_n), \partial_{\boldsymbol{\theta}}\widehat{f}_n(\widehat{\boldsymbol{\theta}}_n)] + O_p(h_{2n})$$
$$+ P\left(\nabla^2 m\{\boldsymbol{Z}, \widehat{f}_n(\widehat{\boldsymbol{\theta}}_n)\}[\partial_{\boldsymbol{\theta}}\widehat{f}_n(\widehat{\boldsymbol{\theta}}_n), \partial_{\boldsymbol{\theta}}\widehat{f}_n(\widehat{\boldsymbol{\theta}}_n)] - \nabla^2 m\{\boldsymbol{Z}, f^*(\widehat{\boldsymbol{\theta}}_n)\}[\partial_{\boldsymbol{\theta}} f^*(\widehat{\boldsymbol{\theta}}_n), \partial_{\boldsymbol{\theta}} f^*(\widehat{\boldsymbol{\theta}}_n)]\right)$$
$$= O_p(n^{-1/4}) + O_p(h_{2n}).$$

Thus, (8) is satisfied as long as $h_{2n} = O_p(n^{-1/4})$.



\end{document}